\documentclass[a4paper,11pt]{article}
\usepackage{graphicx}
\usepackage[dvips]{psfrag}
\usepackage{ulem}
\usepackage{algorithm}
\usepackage{algorithmic}
\usepackage{url}
\usepackage{lscape}
\usepackage{booktabs}
\usepackage{amsthm}
\usepackage{amsmath}
\usepackage{bm}

\usepackage{lscape}
\usepackage{booktabs}
\usepackage{mathtools}
\usepackage{multirow,bigdelim}
\usepackage{colortbl}
\usepackage{setspace}
\usepackage{color}

\makeatletter
\def\section{\@startsection {section}{1}{\z@}{-3.5ex plus -1ex minus -.2ex}{2.3ex plus .2ex}{\large\bf}}

\def\subsection{\@startsection {subsection}{2}{\z@}{-3.5ex plus -1ex minus 
-.2ex}{2.3ex plus .2ex}{\normalsize\bf}}

\def\subsubsection{\@startsection {subsubsection}{3}{\z@}{-3.5ex plus -1ex minus 
-.2ex}{2.3ex plus .2ex}{\normalsize\bf}}
\makeatother

\makeatletter
  \newcounter{subeqncnt}
  \def\thesubeqncnt{\alph{subeqncnt}}%%% ?u????h??õ?v???`??
  \def\subequations{\begingroup%
     \stepcounter{equation}\edef\@tempa{\theequation}%
     \let\c@equation\c@subeqncnt\c@subeqncnt\z@ 
     \edef\theequation{\@tempa\noexpand\thesubeqncnt}}
  
  \makeatother

\makeatletter
\newcount\@cla
\newcount\@clb
\makeatother

\newcommand{\mul}{\multicolumn}

\begin{document}

\title{An Efficient Algorithm for the Escherization Problem in the Polygon Representation}

\author{
Yuichi Nagata\thanks{Corresponding author: nagata@is.tokushima-u.ac.jp}
\and Shinji Imahori\textsuperscript{\dag}
}

\date{}
\maketitle

{
\center 
%\vspace{-10mm}
\footnotesize
\textsuperscript{*} Graduate School of Technology, Industrial and Social Sciences, Tokushima University, 2-1 Minami-jousanjima, Tokushima-shi, Tokushima 770-8506, Japan \\
\textsuperscript{\textsuperscript{\dag}} Department of Information and System Engineering Faculty of Science and Engineering, Chuo University, Bunkyo-ku, Tokyo 112-8551, Japan \\
}
\maketitle

\begin{abstract}	
In the Escherization problem, given a closed figure in a plane, the objective is to find a closed figure that is as close as possible to the input figure and tiles the plane. Koizumi and Sugihara's formulation reduces this problem to an eigenvalue problem in which the tile and input figures are represented as $n$-point polygons. In their formulation, the same number of points are assigned to every tiling edge, which forms a tiling template, to parameterize the tile shape. By considering all possible configurations for the assignment of the $n$ points to the tiling edges, we can achieve much flexibility in terms of the possible tile shapes and the quality of the optimal tile shape improves drastically, at the cost of enormous computational effort. In this paper, we propose an efficient algorithm to find the optimal tile shape for this extended formulation of the Escherization problem.

\bigskip

\noindent {\it Keyword}: Escherization, Escher-like tiling, tessellation, Procrustes distance, isohedral tiling

\end{abstract}

\section{Introduction} \label{sec:intro}

The tiling of a plane is a collection of shapes, called tiles, which cover the plane without any gaps or overlapping. Tiling has attracted much attention, owing to interest in both its practical and mathematical aspects. Tiling theory \cite{kaplan2009introductory} is an elegant branch of mathematics with applications in several areas of computer science. The Dutch artist M.~C.~Escher studied tiling from a mathematical perspective and created many artistic tilings, each of which consists of one or a few recognizable figures, such as animals. Such artistic tiling is now called Escher tiling. People have actively designed new Escher-like tilings, because it is a highly intellectual task to design artistic figures while satisfying the required constraints to realize tiling. Several illustration software packages have been proposed to support the creation process of Escher-like tiling. 

Unlike conventional support systems for Escher-like tilings, Kaplan and Salesin \cite{kaplan2000escherization} developed a method to automatically generate Escher-like tilings. They first formulated an optimization problem called the Escherization problem, which is defined as follows. 

\vspace{2mm}
\noindent {\bf Escherization problem:} Given a closed plane figure $S$ (goal figure), find a closed figure $T$ such that 
\vspace{-1mm}
\begin{enumerate}
\setlength{\parskip}{0cm} % ???
\setlength{\itemsep}{1mm} % ???
\item
$T$ is as close as possible to $S$, and  
\item
copies of $T$ fit together to form a tiling of the plane. 
\end{enumerate}
Their solution method was based on simulated annealing (SA), and more importantly, they proposed a method of parameterizing tile shapes for all isohedral tilings. Here, isohedral tilings are a class of tiling that are not only flexible in modeling tiling patterns, but also mathematically tractable. Their SA succeeded in finding satisfactory tile shapes for convex or nearly convex goal figures. As discussed in their paper, however, their SA algorithm did not perform well for non-convex goal figures. 

Koizumi and Sugihara \cite{koizumi2011maximum} tackled the Escherization problem in a mathematically rigorous way. They represented the tile shape and goal shape as $n$-point polygons. They expressed the constraint conditions imposed on tile shapes for each isohedral type by a homogeneous system of linear equations $A \bm{u} = \bm{0}$, where ${\bm u}$ is a vector representing the coordinates of the tile shape (polygon) and $A$ is a coefficient matrix. The tile shape is parameterized as a general solution of this equation of the form ${\bm u} = B {\bm \xi}$, where $B$ is a matrix consisting of the orthonormal basis of $\mbox{Ker}(A)$ and ${\bm \xi}$ is a parameter vector. They introduced the Procrustes distance \cite{werman1995similarity} to measure the similarity between the two polygons (tile and goal shapes) and showed that the Escherization problem can be reduced to an eigenvalue problem. Koizumi and Sugihara's method performed well for not only convex but also non-convex goal figures. 

As discussed in \cite{koizumi2011maximum} and reported in subsequent studies \cite{imahori2012local,ono2015figure}, a shortcoming of Koizumi and Sugihara's method is its sensitivity to the selection of points for the goal polygon. For a given goal figure, different sets of points can approximate it; however, good results cannot be obtained unless the points are properly arranged on the boundary of the goal figure. Therefore, good sets of points were generated manually through trial and error. To address this problem, Imahori and Sakai \cite{imahori2012local} proposed a local search-based method to find a good set of points for constructing the goal polygon. 

Koizumi and Sugihara's method is interesting, in that the Escherization problem is formalized as an eigenvalue problem that can be solved analytically, despite the abovementioned issue. As we explain later, we can conceptually solve this problem by simply considering all possible configurations for the assignment of the $n$ points to the tiling edges (only one configuration is considered in Koizumi and Sugihara's original formulation), where the tiling edges form a template (e.g., a hexagon) of an isohedral tiling to parameterize the tile shape. However, the required computational complexity increases drastically because we need to solve different eigenvalue problems for each of the possible configurations, whose order reaches $O(n^3)$ and even $O(n^4)$ for certain isohedral types.  

In this paper, we propose an efficient algorithm for an extended version of Koizumi and Sugihara's formulation of the Escherization problem described above. The core ideas are summarized as follows: (i) We need to compute the orthonormal basis of $\mbox{Ker}(A)$ to construct a matrix $B$ for each of the possible configurations for the assignment of $n$ points to the tiling edges (for each isohedral type), which generally takes $O(n^3)$ time. Instead, we propose an $O(n)$ time construction procedure of the matrix $B$. (ii) We propose an $O(n)$ time algorithm to compute the Procrustes distance, which required $O(n^2)$ time in the previous calculation method \cite{imahori2015escher}. (iii) We introduce a mechanism to evaluate a lower bound on the Procrustes distance to omit the calculation of the Procrustes distance that does not lead to an improvement in the best tile shape found so far. 

The remainder of this paper is organized as follows. In Section \ref{sec:2}, we describe Koizumi and Sugihara's formulation of the Escherization problem, as well as related work. In Section \ref{sec:3}, we present an efficient construction procedure of the matrix $B$ (idea (i)) as well as an efficient calculation method of the Procrustes distance (idea (ii)). In Section \ref{sec:4}, we present the overall algorithm, which incorporates a mechanism to calculate a lower bound on the Procrustes distance (idea (iii)). Experimental results are presented in Section \ref{sec:5}, and our conclusions are drawn in Section \ref{sec:6}.

\section{Formulation of the Escherization Problem} \label{sec:2}

In this section, we explain Koizumi and Sugihara's formulation of the Escherization problem, as well as certain related work and concepts.

\subsection{Isohedral tilings}  \label{sec:2_Isohedral}

If all the tiles in a tiling are congruent, we say that the tiling is {\it monohedral}. Among monohedral tilings, if the structure of a tiling can be determined completely by the geometric relationship between a tile and all of its immediate neighbors, we say that the tiling is {\it isohedral}. From this property, any isohedral tiling has a repeating structure of a so-called translation unit (or cluster), which consists of one or several adjacent tiles.  

\begin{figure} [t] %%%%%%%%%%%%%%%%%%%%%%%%%%%%%
\centering
\includegraphics[scale=0.60,keepaspectratio,clip]{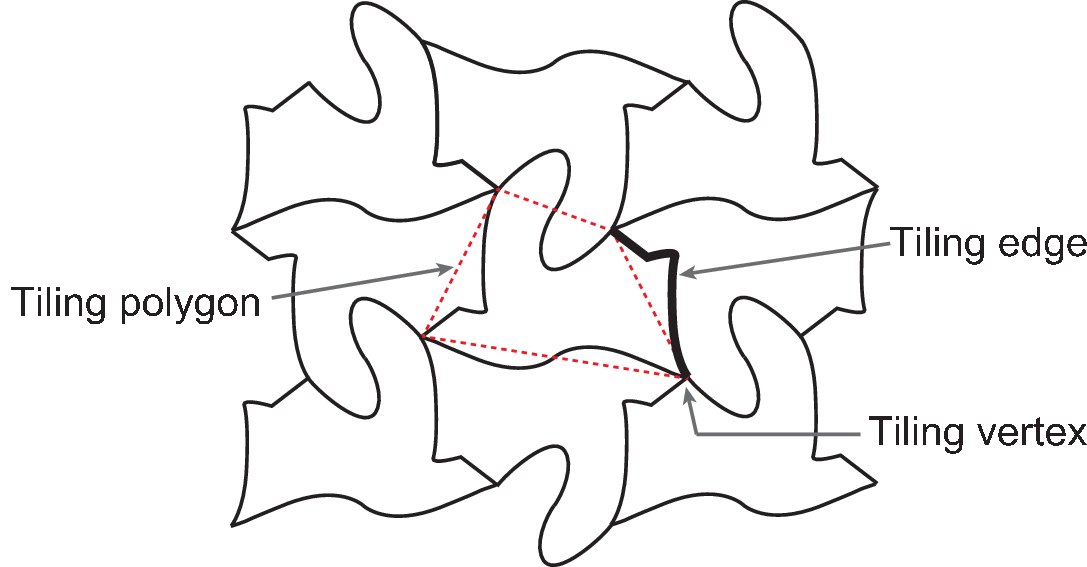}
\caption{Example of an isohedral tiling (IH51).}
\label{fig:tiling_polygon}
\end{figure}

Gr{\"u}nbaum and Shephard \cite{grunbaum1987tilings} classified isohedral tilings into 93 types, which are referred to individually as IH1, IH2, \dots, IH93, based on the adjacency relationship between a tile and its neighbors. Fig. \ref{fig:tiling_polygon} illustrates an example of an isohedral tiling belonging to IH51. For a tile in the tiling, a boundary point in contact with two or more adjacent tiles is called a {\it tiling vertex}, whereas a boundary surface in contact with only one adjacent tile is called a {\it tiling edge}. A tiling polygon is defined as the polygon formed by connecting tiling vertices. 

 %%%%%%%%%%%%%%%%%%%%%%%%%%%%%

\begin{figure} [t] %%%%%%%%%%%%%%%%%%%%%%%%%%%%%
\centering
\includegraphics[scale=0.50,keepaspectratio,clip]{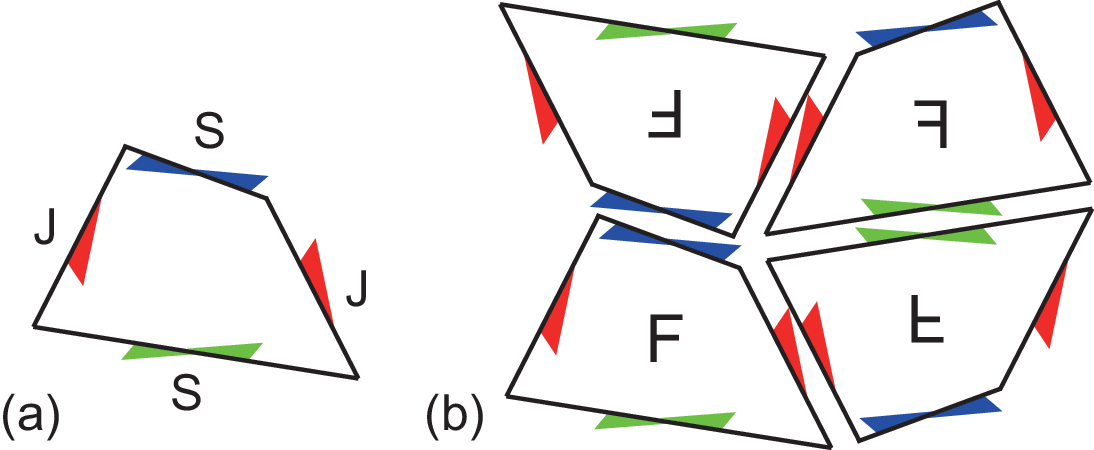}
\caption{(a) Template of IH51 and (b) adjacency relationship between the tiles.}
\label{fig:template}
\end{figure} %%%%%%%%%%%%%%%%%%%%%%%%%%%%%

For each IH type, the nature of the tile shapes can be represented by a template \cite{kaplan2009introductory}. A template represents a tiling polygon, from which all possible tile shapes are obtained by moving the tiling vertices and deforming the tiling edges under the constraints specified by the template. The intuitive representation of the templates of isohedral tilings shown on Tom McLean's website (\url{https://www.jaapsch.net/tilings/mclean/index.html}) is useful. Fig. \ref{fig:template}(a) illustrates a template of IH51, which is given by any quadrilateral consisting of two opposite sides of equal length (indicated by the red arrowheads or J) and two arbitrary sides, meaning that the four tiling vertices on the corners of the quadrilateral are constrained accordingly. The constraints imposed on the tiling edges are represented with colored arrowheads, which take several shapes depending on how they can be deformed. These constraints are closely related to how a tile fits together with its neighbors, and a template also gives information about the adjacency relationship between tiles. Tiles must be placed such that the arrowheads with the same color and shape overlap, as shown in Fig. \ref{fig:template}(b). There are four types of tiling edges, which are named {\bf J}, {\bf S}, {\bf U}, and {\bf I}, depending on the constraints imposed. Two types of arrowheads associated with {\bf J} and {\bf S} edges are shown in Fig. \ref{fig:template}(a). The constraint conditions imposed on the four tiling edge types are described as follows.
\begin{description}
\setlength{\parskip}{0cm} % ???
\setlength{\itemsep}{1mm} % ???
\item[J edge:] 
This edge must be fitted onto another corresponding J edge. This tiling edge can be deformed into any shape, but the corresponding J edge must also be deformed into the same shape, as suggested by the arrowheads. 
\item[S edge:] 
This edge must be fitted onto itself in the opposite direction. From this adjacency relationship, this edge must be symmetric with respect to its midpoint. 
\item[U edge:] 
This edge must be symmetric with respect to a line through its midpoint and orthogonal to it. This constraint condition arises from an internal symmetry of the tile shape, rather than the adjacency relationship.
\item[I edge:] 
This edge must be a straight line. This constraint condition arises from an internal symmetry of the tile shape, rather than the adjacency relationship.
\end{description}
The description of isohedral tilings we have presented is somewhat simplified. For a more rigorous definition, we refer the reader to \cite{kaplan2000escherization,kaplan2009introductory}.

\subsection{Parameterization of isohedral tilings}   \label{sec:2_parameterization}

\begin{figure} [t] %%%%%%%%%%%%%%%%%%%%%%%%%%%%%
\centering
\includegraphics[scale=0.60,keepaspectratio,clip]{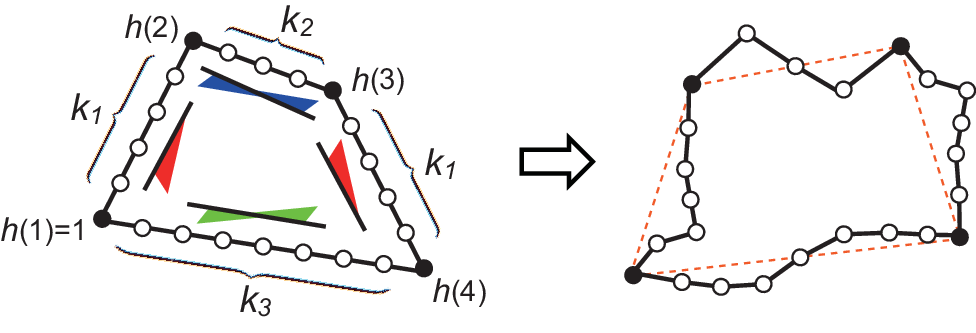}
\caption{Template of IH51 for a specific assignment of the points to the tiling edges (left), and an example of a tile shape (right). }
\label{fig:template_point}
\end{figure}

Koizumi and Sugihara \cite{koizumi2011maximum} modeled the tile shape as an $n$-point polygon. The tile shape (polygon) is constrained to form an isohedral tile, and the constraint conditions depend on the isohedral type. Let us again consider IH51 (see Fig. \ref{fig:template}). To represent the constraint conditions imposed on the $n$ points, it is convenient to define a template as illustrated in Fig. \ref{fig:template_point}. We place the $n$ points on the original template, with a point placed at each of the tiling vertices (represented by black circles) and the remaining points placed on the tiling edges (represented by white circles). This template represents the possible arrangements of the $n$ points, which can be moved under the constraints specified by the template as illustrated in Fig. \ref{fig:template_point}. Koizumi and Sugihara originally placed the same number of points on every tiling edge. Later, Imahori et al. \cite{imahori2015escher} suggested in the conclusion of their paper that it is also possible to place different numbers of points on each tiling edge. We denote the numbers of points placed on the tiling edges (white circles in the figure) as $k_1, k_2, \dots$. For example, these numbers are $k_1=4$, $k_2=3$, and $k_3= 7$ in the example illustrated in Fig. \ref{fig:template_point}.

 %%%%%%%%%%%%%%%%%%%%%%%%%%%%%

Let the $n$ points on the template be indexed clockwise by $1, 2, \dots, n$, starting from one of the tiling vertices. Then, the tile shape is represented as a $2 \times n$ matrix 
\begin{equation}
\label{eq:U}
U = \left(
    \begin{array}{cccc}
      x_1 & x_2 & \dots & x_n \\
      y_1 & y_2 & \dots & y_n 
    \end{array}
  \right),
\end{equation}
where the $i$th column $(x_i, y_i)^{\top}$ is the coordinates of the $i$th point in the $xy$-plane. We also define a $2n$-dimensional vector $\bm{u} = {(x_1, x_2, \dots, x_n, y_1, y_2, \dots, y_n)}^{\top}$. 

We denote the indices of the tiling vertices as $h(s) \ (s =1, \dots, n_v)$, as shown in Fig. \ref{fig:template_point}, where $n_v$ is the number of the tiling vertices and $h(1)=1$. The matrix $U$ is constrained such that the corresponding polygon is consistent with the selected template. For example, if we select IH51 and assign the $n$ points to the tiling edges as specified by $k_1, k_2,$ and $k_3$ (see Fig. \ref{fig:template_point}), the constraint conditions are expressed by the following equation:
\begin{equation}
\label{eq:A_IH51}
\left\{
\begin{array}{llll}
x_{h(1)+i} - x_{h(1)} & = & x_{h(3)+i} - x_{h(3)}    & (i=1, \dots, k_1+1) \\
x_{h(2)+i} - x_{h(2)} & = & -(x_{h(3)-i} - x_{h(3)}) & (i=1, \dots, \lfloor \frac{k_2+1}{2} \rfloor) \\
x_{h(4)+i} - x_{h(4)} & = & -(x_{h(5)-i} - x_{h(1)}) & (i=1, \dots, \lfloor \frac{k_3+1}{2} \rfloor) \\
y_{h(1)+i} - y_{h(1)} & = & -(y_{h(3)+i} - y_{h(3)}) & (i=1, \dots, k_1+1) \\
y_{h(2)+i} - y_{h(2)} & = & -(y_{h(3)-i} - y_{h(3)}) & (i=1, \dots, \lfloor \frac{k_2+1}{2} \rfloor) \\
y_{h(4)+i} - y_{h(4)} & = & -(y_{h(5)-i} - y_{h(1)}) & (i=1, \dots, \lfloor \frac{k_3+1}{2} \rfloor),
\end{array}%
\right.
\end{equation}
where $h(5)$ is defined as $n+1$. Here, we assume that the two J edges form equal and opposite angles with the $x$-axis, which is required to express the constraint conditions as linear equations. This implies that the two opposite J edges are glide-reflection symmetric with respect to the $x$-axis, i.e., one side is obtained by a reflection of the other side with respect to the $x$-axis followed by a translation parallel to the $x$-axis. 

Eq.~(\ref{eq:A_IH51}) is a homogeneous system of linear equations, which can be written as a matrix equation of the form 
\begin{equation}
\label{eq:Au=0}
A \bm{u} = \bm{0},
\end{equation}
where $A$ is an $m' \times 2n$ matrix and $m'$ is the total number of equations in Eq.~(\ref{eq:A_IH51}). The general solution to Eq.~(\ref{eq:Au=0}) is given by
\begin{eqnarray}
\label{eq:u=Bxi}
\bm{u} &=& \xi_1 \bm{b_1} + \xi_2 \bm{b_2} + \cdots + \xi_m \bm{b_m} \nonumber \\
       &=& B \bm{\xi}, 
\end{eqnarray}
where $m$ is the dimension of $\mbox{Ker}(A)$, $\bm{b_i} \ (i=1, \dots, m)$ are the orthonormal basis of $\mbox{Ker}(A)$, and $\xi_i \ (i=1, \dots, m)$ are the parameters. The latter form is a matrix representation with the $2n \times m$ matrix $B=(\bm{b_1} \ \bm{b_2} \ \dots \ \bm{b_m})$ and vector $\bm{\xi} = (\xi_1, \xi_2, \dots, \xi_m)^{\top}$. 

Tile shapes for every isohedral type can be parameterized in the form of Eq.~(\ref{eq:u=Bxi}), where the matrix $B$ depends on the isohedral type as well as the assignment of the $n$ points to the tiling edges. Because the order of $m$ is $O(n)$, it takes $O(n^3)$ time to compute the orthonormal basis of $\mbox{Ker}(A)$ by using an appropriate calculation method (e.g., Gaussian elimination followed by the Gram--Schmidt orthogonalization).

\subsection{Similarity measure}   \label{sec:2_similarity}

The goal figure is represented as a polygon with $n$ points located on its boundary. Let the $n$ points be indexed clockwise by $1, 2, \dots, n$ and their coordinates be represented by a $2 \times n$ matrix 
\begin{equation}
\label{eq:W}
W = \left(
    \begin{array}{cccc}
      x^w_1 & x^w_2 & \dots & x^w_n \\
      y^w_1 & y^w_2 & \dots & y^w_n 
    \end{array}
  \right),
\end{equation}
where the $i$th column $(x^w_i, y^w_i)^{\top}$ is the coordinates of the $i$th point. We also define a $2n$-dimensional vector $\bm{w} = {(x^w_1, x^w_2, \dots, x^w_n, y^w_1, y^w_2, \dots, y^w_n)}^{\top}$ and $n$-dimensional vectors $\bm{w_x} = {(x^w_1, x^w_2, \dots, x^w_n)}^{\top}$ and $\bm{w_y} = {(y^w_1, y^w_2, \dots, x^y_n)}^{\top}$.

Koizumi and Sugihara employed the Procrustes distance \cite{werman1995similarity} to measure the similarity between the two polygons $U$ and $W$. Let $\|X\|$ be the Frobenius norm of a matrix $X$. The square of the Procrustes distance, which we denote as $d^2(U,W)$, is defined and transformed as follows (see \cite{werman1995similarity} for more details):
\begin{equation}
\label{eq:procrustes}
d^2(U,W) = \min_{s, \theta} {\left\| sR(\theta)\frac{U}{\|U\|} - \frac{W}{\|W\|}  \right\|}^2 = 1 - \frac{ {\|UW^{\top}\|}^2 + 2\det(U{W}^{\top}) }{{\|U\|}^2 {\|W\|}^2},  
\end{equation}
where $s$ is a scalar expressing expansion (or contraction) and $R(\theta)$ is the rotation matrix by angle $\theta$. From the definition, the Procrustes distance is scale- and rotation-invariant. The property of rotation-invariance is indispensable\footnote{The property of scale-invariance is not necessary here. Therefore, a distance measure defined as $\min_{\theta} {\left\| R(\theta) U - W  \right\|}^2$ gives the same result, except for the size.}, because the parameterized tile shape $U$ for certain isohedral types such as IH51 can only appear in a specific orientation.

Minimizing $d^2(U,W)$ is equivalent to maximizing 
\begin{equation}
\label{eq:procrustes_part}
\frac{ {\|UW^{\top}\|}^2 + 2\det(U{W}^{\top}) }{{\|U\|}^2}. 
\end{equation}
Koizumi and Sugihara showed that Eq.~(\ref{eq:procrustes_part}) could be simplified to
\begin{equation}
\label{eq:uVu}
\frac{{\bm{u}}^{\top} V \bm{u}} {{\bm{u}}^{\top}\bm{u}}, 
\end{equation}
where $V$ is the $2n \times 2n$ symmetric matrix defined by 
\begin{equation}
\label{eq:V}
V = \left(
    \renewcommand{\arraystretch}{1.1}
    \begin{array}{cc}
      \bm{w_x}{\bm{w_x}}^{\top} + \bm{w_y}{\bm{w_y}}^{\top} & \ \ \bm{w_x}{\bm{w_y}}^{\top} - \bm{w_y}{\bm{w_x}}^{\top} \\ 
      \bm{w_y}{\bm{w_x}}^{\top} - \bm{w_x}{\bm{w_y}}^{\top} & \ \ \bm{w_x}{\bm{w_x}}^{\top} + \bm{w_y}{\bm{w_y}}^{\top} 
    \end{array}
  \right).
\end{equation}

We should note that, when calculating the Procrustes distance between two polygons $U$ and $W$, we need to consider $n$ different numbering schemes for the goal polygon $W$ by shifting the first point of the numbering. Therefore, we define $W_j \ (j=1, 2, \dots, n)$ as the $2 \times n$ matrix obtained from $W$ by renumbering the index such that the $j$th point of the original index becomes the first point of the new index. We further define $\bm{w_j}$ and $V_j$ for $\bm{w}$ and $V$, respectively, in the same manner.

\subsection{Escherization problem as an eigenvalue problem}  \label{sec:2_optimization}

From Eqs.~(\ref{eq:u=Bxi}) and (\ref{eq:uVu}), and the relation $B^{\top} B = I$ (identity matrix), the Escherization problem can be formulated as the following unconstrained optimization problem:
\begin{equation}
\label{eq:xiBVBxi}
\mbox{maximize:} \ \dfrac{{\bm{\xi}}^{\top} B^{\top} V B \bm{\xi}} {{\bm{\xi}}^{\top}\bm{\xi}}.
\end{equation}

The optimization problem (\ref{eq:xiBVBxi}) is known as the Rayleigh quotient; the optimal value is the maximum eigenvalue of $B^{\top}VB$ and the optimal solution ${\bm \xi^*}$ is the eigenvector associated with the maximum eigenvalue (the length is arbitrary). By utilizing the fact that $B^{\top}VB$ is a symmetric positive semidefinite matrix of maximum rank two, Imahori et al. \cite {imahori2015escher} proposed a projection method \cite{saad2011numerical} to compute the maximum eigenvalue and its corresponding eigenvector in $O(n^2)$ time. 

To summarize Eqs.~(\ref{eq:procrustes})--(\ref{eq:xiBVBxi}), we have the following equation: 
\begin{equation}
\label{eq:d2_procrustes}
\min_{\substack{{\bm u} = B {\bm \xi} \\ {\bm \xi \in R^m} } } d^2(U,W) = 1 - \frac{\lambda}{{\|W\|}^2},
\end{equation}
where $\lambda$ is the maximum eigenvalue of ${B}^{\top}V B$. As already stated, the matrix $B$ depends on the isohedral type as well as the assignment of $n$ points to the tiling edges. In Koizumi and Sugihara's original formulation \cite{koizumi2011maximum}, for each isohedral type IH$i \ (i = 1, 2, \dots, 93)$, the same number of points were assigned to every tiling edge (i.e., only one configuration was considered) and we denote the corresponding matrix $B$ as $B_i$. Recall that the matrix $V$ depends on the first point (denoted by $j$) for the numbering of the goal polygon. Consequently, we need to compute  
\begin{equation}
\label{eq:d2_procrustes_ij}
\min_{\substack{{\bm u} = B_i {\bm \xi} \\ {\bm \xi \in R^m} } } d^2(U,W_j) = 1 - \frac{\lambda_{ij}}{{\|W\|}^2},
\end{equation}
for all combinations of $i \ (= 1, 2, \dots, 93)$ and $j \ (=1, 2, \dots, n)$, where $\lambda_{ij}$ is the maximum eigenvalue of $B_i^{\top}V_j B_i$. Then, we select the best tile shape as that with the minimum distance value.

As suggested by Imahori et al. \cite{imahori2015escher}, we need to consider only 28 isohedral types of those originally classified by Heesch and Kienzle \cite{heesch1963}, because the remaining 65 isohedral types are regarded as special cases of the 28 types; other 65 types are obtained by imposing internal symmetries on the 28 types. For each isohedral type IH$i$, it takes $O(n^3)$ time to compute Eq~(\ref{eq:d2_procrustes_ij}) for $j=1, 2, \dots, n$, because it takes $O(n^3)$ time to compute $B_{i}$ and $O(n^2)$ time to compute the maximum eigenvalue of $B_{i}^{\top} V_j B_{i}$ (for each value of $j$).
                                                                                                                         
\section{Exhaustive Search for the Escherization Problem and Efficient Calculation Methods}  \label{sec:3}

In this section, we explain the optimization problem to be solved and present efficient calculation methods for solving this problem.  

\subsection{Motivation} \label{sec:3_Motivation}

Koizumi and Sugihara \cite{koizumi2011maximum} assigned the same number of points to every tiling edge of the template for each isohedral type. However, when the $n$ points are located at equal intervals on the boundary of the goal figure to form a goal polygon $W$, the optimal tile shape $U$ consists of tiling edges of nearly equal length, even though the most satisfactory tile shape consists of tiling edges with substantially different lengths. The only way to alleviate this problem would be to place the points non-uniformly on the boundary of the goal figure to form a goal polygon, as long as the same number of points are assigned to every tiling edge of the template. In fact, satisfactory results cannot be obtained unless the points are properly arranged on the boundary of the goal figure, and they manually generated a good set of points through trial and error. To address this problem, Imahori and Sakai \cite{imahori2012local} proposed a local search-based method to find a good set of points for the goal polygon. 

A more straightforward approach to overcome the weakness of Koizumi and Sugihara's method would be to assign different numbers of points to the tiling edges and to form the goal polygon by placing $n$ points at equal intervals on the boundary of the goal figure. As already stated in Section \ref{sec:2_parameterization}, this extension is feasible and would create a great deal of flexibility in the possible tile shapes. However, it seems to be computationally unrealistic to consider all possible configurations for the assignment of $n$ points to the tiling edges (the details are described in the next subsection). Therefore, our motivation is to develop an efficient algorithm to compute the optimal tile shape for this problem.

\subsection{Exhaustive Koizumi and Sugihara's formulation and basic ideas} \label{sec:3_Exhaustive}

Let $I$ be a set of the indices for the isohedral types considered for the optimization and $K_i$ be a set of all possible configurations for the assignment of $n$ points to the tiling edges for an isohedral type $i \in I$ ($i \in I $ indicates IH$i$). For example, $K_{51}=\{(k_1, k_2) \mid 0 \leq k_1, 0 \leq k_2, 2 k_1 + k_2 \leq n-4 \}$, with $k_3$ determined by $k_3=n-4-2 k_1 - k_2$ (see Fig. \ref{fig:template_point}). Recall that the matrix $B$ depends on $i \in I$ and $k \in K_i$, and we denote it as $B_{ik}$. Let $J = \{1, 2, \dots, n\}$ be a set of the indices of the first point for the $n$ different numbering schemes of the goal polygon. We define an exhaustive version of Koizumi and Sugihara's formulation of the Escherization problem as the search for the tile shape $U$ closest to the goal shape $W$ with respect to the Procrustes distance among all combinations of $i \in I$, $k \in K_i$, and $j \in J$. To find the optimal tile shape for this problem, we need to compute 
\begin{equation}
\label{eq:eval_ikj}
eval_{ikj} = \min_{\substack{{\bm u} = B_{ik} {\bm \xi} \\ {\bm \xi \in R^m} } } d^2(U,W_j) = 1 - \frac{\lambda_{ikj}}{{\|W\|}^2}
\end{equation}
for all combinations of $i \in I$, $k \in K_i$, and $j \in J$, where $\lambda_{ikj}$ is the maximum eigenvalue of $B_{ik}^{\top} V_j B_{ik}$.

\begin{figure} [t] %%%%%%%%%%%%%%%%%%%%%%%%%%%%%
\centering
\includegraphics[scale=0.55,keepaspectratio,clip]{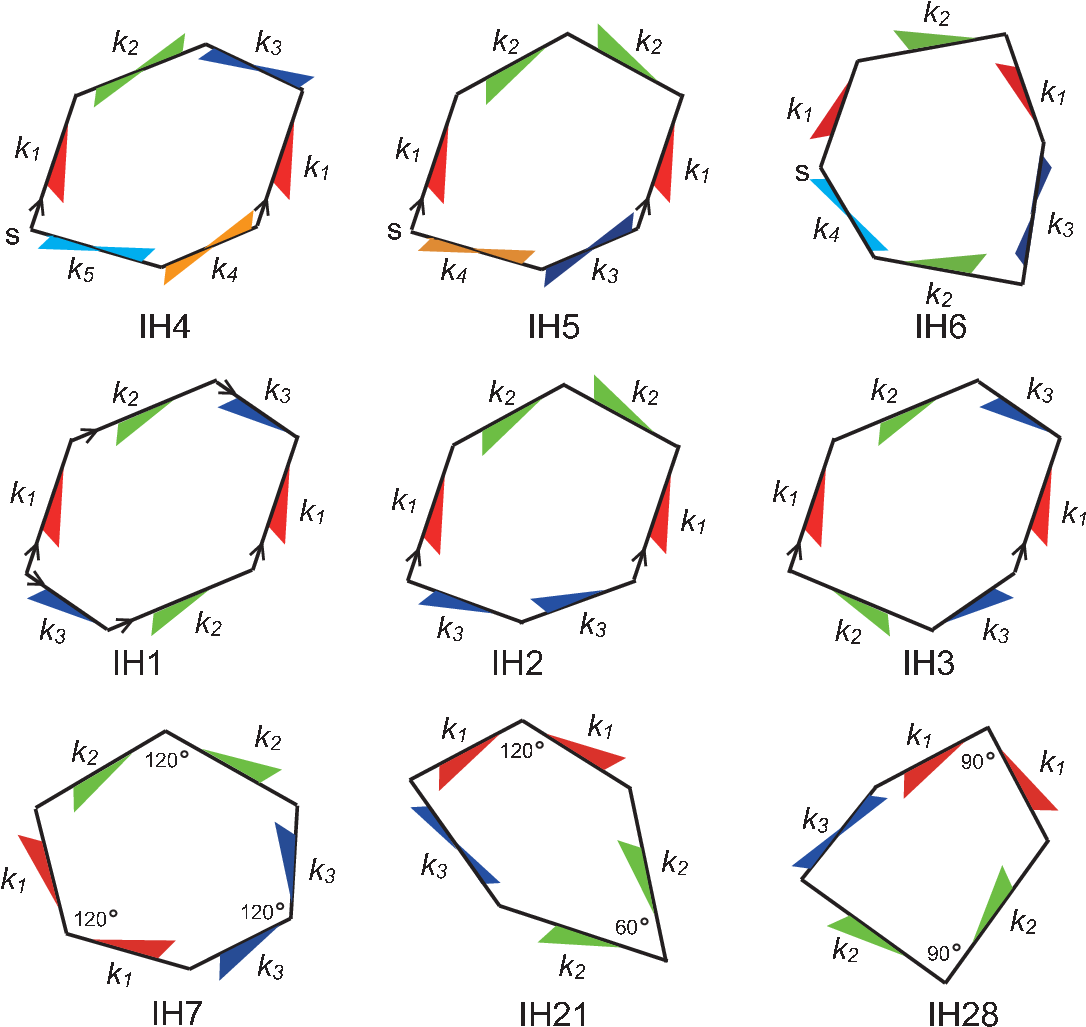}
\caption{Templates of the nine most general isohedral types. The first tiling vertex is marked with ``s'' (for IH4, IH5, and IH6). Two opposite J edges marked with $\wedge$ are parallel to one another. Consecutive J edges must form a specified angle if designated. Otherwise, each of the pairs of J edges is glide-reflection symmetric with respect to the $x$-axis or $y$-axis.}
\label{fig:IH_all}
\end{figure}

As stated in Section \ref{sec:2_optimization}, we need to consider 28 isohedral types for the optimization. In our model, if the template of an isohedral type is obtained by removing one or more tiling edges from the template of any of these 28 isohedral types, we exclude such isohedral type because we can approximately remove a tiling edge by assigning no point to it (but a small tiling edge consisting of two tiling vertices still remains). As described in \cite{Schattschneider2004}, these 28 isohedral types are obtained in this way from the nine most general isohedral types. Fig. \ref{fig:IH_all} illustrates the templates of the nine most general isohedral types and we consider only these types for the optimization. 

 %%%%%%%%%%%%%%%%%%%%%%%%%%%%%

Algorithm \ref{alg:exhaustive_search} depicts the exhaustive search algorithm for the exhaustive Koizumi and Sugihara's formulation of the Escherization problem. In this algorithm, the minimum value of $eval_{ikj}$ found so far is stored in $eval_{min}$ and the corresponding best tile shape is represented as $\bm{u^*}$. By using Koizumi and Sugihara's algorithm in concert with the projection method developed by Imahori et al. \cite{imahori2015escher}, for each $i \in I$ and $k \in K_i$, it takes $O(n^3)$ time to perform the procedure (lines 4--14), because it takes $O(n^3)$ time to compute $B_{ik}$ (line 4) and $O(n^2)$ time to compute the maximum eigenvalue of $B_{ik}^{\top} V_j B_{ik}$ (line 7). Because the order of $K_i$ reaches $O(n^3)$ for IH5 and IH6, and $O(n^4)$ for IH4, executing this algorithm based on the previous approaches takes very long computational times. To execute Algorithm \ref{alg:exhaustive_search} more efficiently, we propose three techniques, as described in the following.
\begin{enumerate}
\item
We propose an efficient calculation method of the matrix $B_{ik}$ (Section \ref{sec:3_B}). It takes $O(n)$ time, whereas this calculation in the previous approaches took $O(n^3)$ time \cite{koizumi2011maximum,imahori2012local,imahori2015escher}.
\item
We propose an efficient calculation method of the Procrustes distance (Section \ref{sec:3_Procrustes}). It takes $O(n)$ time, whereas the projection method developed by Imahori et al. \cite{imahori2015escher} took $O(n^2)$ time. 
\item
We propose an efficient algorithm that incorporates a mechanism to evaluate a lower bound on the value of $eval_{ikj}$ in the course of the search, in order to eliminate unnecessary computations of $eval_{ikj}$ that do not yield an improvement in $eval_{mim}$ (Section \ref{sec:4}).  
\end{enumerate}

\begin{algorithm}
\caption{: {\sc Exhaustive Search}}
\label{alg:exhaustive_search}
\algsetup{
linenosize=\small,
linenodelimiter=:
}
\begin{algorithmic}[1]
\STATE $eval_{min} = \infty$;
\FOR{$i \in I$ (nine isohedral types)}
\FOR{$k \in K_i$ (possible assignments of the $n$ points to the tiling edges)}
\STATE Compute $B_{ik}$;
\FOR{$j \in J$ (choices of the first point for the goal polygon)}
\STATE Compute $V_j$;
\STATE Compute the maximum eigenvalue of $B_{ik}^{\top} V_j B_{ik}$ $\rightarrow \lambda_{ikj}$;
\STATE Compute $eval_{ikj}$ from $\lambda_{ikj}$ (Eq.~(\ref{eq:eval_ikj}) or (\ref{eq:eval_ikj_Euclid}));
\IF{$eval_{ikj} < eval_{min}$}
\STATE $eval_{min} := eval_{ikj}$;
\STATE Compute the maximum eigenvector of $B_{ik}^{\top} V_j B_{ik}$ $\rightarrow \bm{\xi}$;
\STATE Compute $\bm{u^*}=B_{ik} \bm{\xi}$;
\ENDIF
\ENDFOR
\ENDFOR
\ENDFOR
\RETURN $\bm{u^*}$;
\end{algorithmic}
\end{algorithm}

\subsection{Efficient computation for the matrix $B_{ik}$} \label{sec:3_B}

When the constraint conditions imposed on the tile shape $U$ are expressed in the form of Eq.~(\ref{eq:Au=0}), it generally takes $O(n^3)$ time to compute the matrix $B_{ik}$. We propose an $O(n)$ time algorithm for the construction of $B_{ik}$. We first explain how to construct $B_{ik}$ for the isohedral type IH51 (see Fig. \ref{fig:template_point}) as an easily understood example. In fact, we can construct $B_{ik}$ in a similar manner for any $i \in I$ and $k \in K_i$, as we will explain at the end of this subsection. 

A basic idea is to divide the tiling vertex parameterization and tiling edge parameterization more explicitly as Kaplan and Salesin did \cite{kaplan2000escherization} (see also \cite{kaplan2009introductory}). In fact, our parametrization method up to Eq.~(\ref{eq:u=Bd}) would be essentially the same as their method, except for some details. Let $n_v$ be the number of the tiling vertices, and we define the $2n_v$-dimensional vector $\bm{u_v} = {(x^v_1, x^v_2, \dots, x^v_{n_v}, y^v_1, y^v_2, \dots, y^v_{n_v})}^{\top}$ to represent the coordinates of the tiling vertices. Then, the constraint conditions imposed on the tiling vertices can be expressed in the same way as in Eq.~(\ref{eq:A_IH51}). These constraint conditions for IH51 are expressed by
\begin{equation}
\label{eq:constraint_vertex}
\left\{
\begin{array}{lll}
x^v_{2} - x^v_{1} & = & x^v_{4} - x^v_{3}  \\
y^v_{2} - y^v_{1} & = & -(y^v_{4} - y^v_{3})  
\end{array}%
\right.
.
\end{equation}
Then, $\bm{u_v}$ is parameterized by 
\begin{equation}
\label{eq:uv}
\bm{u_v} = B_v \bm{\xi_d} 
\end{equation}
in the same way as in Eq.~(\ref{eq:u=Bxi}), where $B_v$ is the $2n_v \times m_d$ matrix and $\bm{\xi_d}$ is an $m_d$-dimensional vector ($m_d$ depends on the isohedral type). Let ${\bm{d_l}}^{\top}$ denote the $l$th row of $B_v$ and $B_v$ be represented by 
\begin{equation}
\label{eq:Bv}
B_v =
\begin{pmatrix}
{\bm{d_1}}^{\top} \\
{\bm{d_2}}^{\top} \\
\vdots \\
{\bm{d_{2n_v}}}^{\top} \\
\end{pmatrix}
.
\end{equation}

Next, we explain how the tile shape $U$ (equivalently, $\bm{u}$) is parameterized. Considering the properties of J and S edges, tile shapes of IH51 with a given assignment of $n$ points to the tiling edges $(k_1, k_2) \in K_{51}$ can be parameterized directly as follows:
%\vspace{3mm}
\begin{equation}
%\footnotesize
\thickmuskip=0mu
\medmuskip=0mu
\thinmuskip=0mu
\label{eq:parameterize_u_proposed}
\renewcommand{\arraystretch}{1.20} % 1.28
\begin{bmatrix*}[c]
\colorbox[gray]{.9}{$x_{h(1)}$} \\
x_{h(1)+1} \\
\scalebox{1.0}[0.8]{\vdots} \\
x_{h(1)+k_1} \\
\colorbox[gray]{.9}{$x_{h(2)}$} \\
x_{h(2)+1} \\
\scalebox{1.0}[0.8]{\vdots} \\
\vspace{-1mm}
x_{h(2)+\lfloor \frac{k_2}{2} \rfloor} \\
\vspace{-1mm}
x_{h(2)+\lfloor \frac{k_2+1}{2} \rfloor} \\
x_{h(3)-\lfloor \frac{k_2}{2} \rfloor} \\
\scalebox{1.0}[0.8]{\vdots} \\
x_{h(3)-1} \\
\colorbox[gray]{.9}{$x_{h(3)}$} \\
x_{h(3)+1} \\
\scalebox{1.0}[0.8]{\vdots} \\
x_{h(3)+k_1} \\
\colorbox[gray]{.9}{$x_{h(4)}$} \\
\scalebox{1.0}[0.8]{\vdots} \\
\end{bmatrix*} 
=
\renewcommand{\arraystretch}{1.22} % 1.32
\begin{bmatrix*}[c]
\colorbox[gray]{.9}{$\bm{d_1}^{\top}$} \\
\bm{d_1}^{\top} \\
\scalebox{1.0}[0.8]{\vdots} \\
\bm{d_1}^{\top} \\
\colorbox[gray]{.9}{$\bm{d_2}^{\top}$} \\
\frac{1}{2}(\bm{d_2} \!+ \! \bm{d_3})^{\top}\\
\scalebox{1.0}[0.8]{\vdots} \\
\frac{1}{2}(\bm{d_2} \!+ \! \bm{d_3})^{\top}\\
\frac{1}{2}(\bm{d_2} \!+ \! \bm{d_3})^{\top}\\
\frac{1}{2}(\bm{d_2} \!+ \! \bm{d_3})^{\top}\\
\scalebox{1.0}[0.8]{\vdots} \\
\frac{1}{2}(\bm{d_2} \!+ \! \bm{d_3})^{\top}\\
\colorbox[gray]{.9}{$\bm{d_3}^{\top}$} \\
\bm{d_3}^{\top} \\
\scalebox{1.0}[0.8]{\vdots} \\
\bm{d_3}^{\top} \\
\colorbox[gray]{.9}{$\bm{d_4}^{\top}$} \\
\scalebox{1.0}[0.8]{\vdots} \\
\end{bmatrix*}
\bm{\xi_d}
+
\renewcommand{\arraystretch}{1.28} % 1.32
\tfrac{\xi_1^s}{\sqrt{2}}
\begin{bmatrix*}[c]
\colorbox[gray]{.9}{$\color[gray]{0.9}{{}^1 }$} \\
1 \\
\\
\\
\colorbox[gray]{.9}{$\color[gray]{0.9}{{}^1 }$} \\
\\
\\
\\
\\
\\
\\
\\
\colorbox[gray]{.9}{$\color[gray]{0.9}{{}^1 }$} \\
1 \\
\\
\\
\colorbox[gray]{.9}{$\color[gray]{0.9}{{}^1 }$} \\
\scalebox{1.0}[0.8]{\vdots} \\
\end{bmatrix*}
+
\cdots
+
\tfrac{\xi_{k_1}^s}{\sqrt{2}}
\begin{bmatrix*}[c]
\colorbox[gray]{.9}{$\color[gray]{0.9}{{}^1 }$} \\
\\
\\
1 \\
\colorbox[gray]{.9}{$\color[gray]{0.9}{{}^1 }$} \\
\\
\\
\\
\\
\\
\\
\\
\colorbox[gray]{.9}{$\color[gray]{0.9}{{}^1 }$} \\
\\
\\
1 \\
\colorbox[gray]{.9}{$\color[gray]{0.9}{{}^1 }$} \\
\scalebox{1.0}[0.8]{\vdots} \\
\end{bmatrix*}
+
\tfrac{\xi_{k_1+1}^s}{\sqrt{2}}
\begin{bmatrix*}[c]
\colorbox[gray]{.9}{$\color[gray]{0.9}{{}^1 }$} \\
\\
\\
\\
\colorbox[gray]{.9}{$\color[gray]{0.9}{{}^1 }$} \\
1\\
\\
\\
\\
\\
\\
\scalebox{0.9}[1]{-1}\\
\colorbox[gray]{.9}{$\color[gray]{0.9}{{}^1 }$} \\
\\
\\
\\
\colorbox[gray]{.9}{$\color[gray]{0.9}{{}^1 }$} \\
\scalebox{1.0}[0.8]{\vdots} \\
\end{bmatrix*}
+
\cdots
+
\tfrac{\xi_{k_1+ \lfloor \frac{k_2}{2} \rfloor }^s}{\sqrt{2}}
\begin{bmatrix*}[c]
\colorbox[gray]{.9}{$\color[gray]{0.9}{{}^1 }$} \\
\\
\\
\\
\colorbox[gray]{.9}{$\color[gray]{0.9}{{}^1 }$} \\
\\
\\
1 \\
\\
1 \\
\\
\\
\colorbox[gray]{.9}{$\color[gray]{0.9}{{}^1 }$} \\
\\
\\
\\
\colorbox[gray]{.9}{$\color[gray]{0.9}{{}^1 }$} \\
\scalebox{1.0}[0.8]{\vdots} \\
\end{bmatrix*}
%+
\cdots
,
%\vspace{3mm}
\end{equation} 
where the blank elements in the column vectors are zero and the rows that parameterize the coordinates of the tiling vertices are shaded. From Eqs.~(\ref{eq:uv}) and (\ref{eq:Bv}), the $xy$-coordinates of the tiling vertices $(x_{h(s)}, y_{h(s)})$ $(s=1,2, \dots, n_v)$ are parameterized with $({\bm{d_{s}}}^{\top} \bm{\xi_d}, {\bm{d_{n_v+s}}}^{\top} \bm{\xi_d})$ in the above formula. The coordinates of the other points are expressed as the difference from those of the tiling vertices (or the midpoint of the two tiling vertices) on the same tiling edge. For example, according to the constraint conditions imposed on the two opposite J edges, the $x$-coordinates of the points $h(1)+i$ and $h(3)+i$ $(i=1, 2, \dots, k_1)$ are parameterized with the parameters $\xi^s_1, \dots, \xi^s_{k_1}$ and the column vectors ($\frac{1}{\sqrt{2}}$ is factored out) multiplied by these parameters, as shown in Eq.~(\ref{eq:parameterize_u_proposed}). In the same way, according to the constraint conditions imposed on the S edge starting from index $h(2)$, the $x$-coordinates of the points $h(2)+i$ and $h(3)-i$ $(i=1, \dots, \lfloor \frac{k_2}{2} \rfloor)$ are parameterized with the parameters $\xi^s_{k_1+1}, \dots, \xi^s_{k_1+ \lfloor \frac{k_2}{2} \rfloor}$ and the column vectors ($\frac{1}{\sqrt{2}}$ is factored out) multiplied by these parameters, as shown in Eq.~(\ref{eq:parameterize_u_proposed}). We should note that if $k_2$ is an even number ($\lfloor \frac{k_2}{2} \rfloor = \lfloor \frac{k_2+1}{2} \rfloor$), the row that parameterizes $x_{h(2)+\lfloor \frac{k_2+1}{2} \rfloor}$ must be removed from Eq.~(\ref{eq:parameterize_u_proposed}). On the contrary, if $k_2$ is an odd number ($\lfloor \frac{k_2}{2} \rfloor + 1 = \lfloor \frac{k_2+1}{2} \rfloor$), the point $h(2) + \lfloor \frac{k_2+1}{2} \rfloor \ (= h(3) - \lfloor \frac{k_2+1}{2} \rfloor)$ is the midpoint between the two tiling vertices $h(2)$ and $h(3)$, and the $x$-coordinate of the midpoint is therefore parameterized as $\frac{1}{2}{({\bm{d_{2}}} + {\bm{d_{3}}})}^{\top} \bm{\xi_d}$, without an additional parameter.        

Next, we express Eq.~(\ref{eq:parameterize_u_proposed}) in a matrix form. Let the $2n \times m_d$ matrix multiplied by the vector $\bm{\xi_d}$ on the right-hand side of Eq.~(\ref{eq:parameterize_u_proposed}) be denoted as $B_d$. In the same way, let the $2n$-dimensional column vectors multiplied by the parameters $\xi_i$ $(i = 1, 2, \dots, m_s)$ be denoted as $\bm{b^s_i}$ $(i = 1, 2, \dots, m_s)$, where $m_s$ depends on $k \in K_i$. We express $B_d$ and define the $2n \times m_s$ matrix $B_s$ and $m_s$-dimensional column vector $\bm{\xi_s}$ as follows:
\begin{eqnarray}
\label{eq:Bd}
\begin{aligned}
B_d &=& \left(\bm{b^d_1} \ \bm{b^d_2} \ \cdots \ \bm{b^d_{m_d}}\right) \\
B_s &=& \left(\bm{b^s_1} \ \bm{b^s_2} \ \cdots \ \bm{b^s_{m_s}}\right) \\
\bm{\xi_s} &=& {(\xi^s_1 \ \xi^s_2 \ \cdots \ \xi^s_{m_s})}^{\top},
\end{aligned}
\end{eqnarray}
where $\bm{b^d_i}$ is the $i$th column of $B_d$. Then, a matrix representation of Eq.~(\ref{eq:parameterize_u_proposed}) is given by 
\begin{equation}
\label{eq:u=Bd}
\bm{u} = B_d \bm{\xi_d} + B_s \bm{\xi_s} = (B_d \  B_s) 
\begin{pmatrix*}[c] 
\bm{\xi_d} \\
\bm{\xi_s} \\
\end{pmatrix*}
.
\end{equation}

Some desirable properties of this parameterization are summarized as follows:
\begin{itemize}
\item
The matrices $B_d$ and $B_s$ are obtained directly, without the process of transforming the matrix $A$ into $B$ (see Eqs.~(\ref{eq:Au=0}) and (\ref{eq:u=Bxi})), once $B_v$ is computed (Eq.~(\ref{eq:Bv})).
\item
$B_s$ is a sparse matrix.  
\item
All column vectors of $B_s$ are mutually orthogonal.
\end{itemize}
Note that the last property clearly holds in Eq.~(\ref{eq:parameterize_u_proposed}), because nonzero elements in the column vectors $\bm{b^s_1}, \bm{b^s_2}, \dots, \bm{b^s_{m_s}}$ never overlap with each other. The column vectors of $B_d$ are not, however, orthogonal to each other and they are also not orthogonal to the column vectors of $B_s$. Therefore, we linearly transform $\bm{{b}^d_1}, \bm{{b}^d_2}, \dots, \bm{{b}^d_{m_d}}$ into a set of column vectors $\bm{{b''}^d_1}, \bm{{b''}^d_2}, \dots, \bm{{b''}^d_{m_d}}$, such that $\bm{{b''}^d_1}, \bm{{b''}^d_2}, \dots, \bm{{b''}^d_{m_d}}, \bm{b^s_1}, \bm{b^s_2}, \dots, \bm{b^s_{m_s}}$ becomes an orthonormal basis of $\mbox{span}(\bm{{b}^d_1}, \bm{{b}^d_2}, \dots, \bm{{b}^d_{m_d}}, \bm{b^s_1}, \bm{b^s_2}, \dots, \bm{b^s_{m_s}})$.

We first consider the orthogonalization of $\bm{b^d_i} \ (i = 1, \dots, m_d)$ against $\bm{b^s_j} \ (j = 1, \dots, m_s)$. This can be achieved simply by using the Gram--Schmidt method. Let $\bm{{b'}^d_i} \ (i = 1, 2, \dots, m_d)$ be a column vector obtained from $\bm{b^d_i}$ after applying the Gram--Schmidt method, as follows: 
\begin{equation}
\label{eq:gram_schmidt}
\bm{{b'}^d_i} = \bm{b^d_i} - \langle \bm{b^d_i}, \bm{b^s_1}\rangle  \bm{b^s_1} - \langle \bm{b^d_i}, \bm{b^s_2}\rangle  \bm{b^s_2} - \cdots -  \langle \bm{b^d_i}, \bm{b^s_{m_s}}\rangle  \bm{b^s_{m_s}}, 
\end{equation}
where $\langle \bm{x}, \bm{y}\rangle $ is the inner product of the two vectors.  

Let the $2n \times m_d$ matrix ${B'}_d$ be defined by
\begin{equation}
\label{eq:B'd}
{B'}_d = \left(\bm{{b'}^d_1} \ \bm{{b'}^d_2} \ \cdots \ \bm{{b'}^d_{m_d}} \right).
\end{equation}
Here, the important point is that the rows of ${B'}_d$ take only a limited number of vector values. From the form of Eq.~(\ref{eq:parameterize_u_proposed}), the vectors $\bm{b^s_j} \ (j = k_1+1, \dots, k_1+\lfloor \frac{k_2}{2} \rfloor)$, which parameterize the $x$-coordinates of the S edge, are already orthogonal to the vectors $\bm{b^d_i} \ (i = 1, \dots, m_d)$. Therefore, the $l$th $(h(2) < l < h(3))$ rows of ${B}_d$ remain the same after applying the Gram--Schmidt method. However, the vectors $\bm{b^s_j} \ (j = 1, \dots, k_1) $, which parameterize the $x$-coordinates of the two J edges, are not orthogonal to the vectors $\bm{b^d_i} \ (i = 1, \dots, m_d)$. From the form of Eq.~(\ref{eq:parameterize_u_proposed}), every pair of the ($h(1)+i$)th and ($h(3)+i$)th rows of ${B}_d$ becomes the same pair of two row vectors after applying the Gram--Schmidt method, regardless of the value of $i = 1, 2, \dots, k_1$. The two row vectors, which we denote by ${\bm{d^f_1}}^{\top}$ and ${\bm{d^f_3}}^{\top}$, are obtained by 
$
\begin{pmatrix*}[c]
{\bm{d^f_1}}^{\top} \\ 
{\bm{d^f_3}}^{\top}
\end{pmatrix*}  
= 
\begin{pmatrix*}[c]
{\bm{d_1}}^{\top} \\ 
{\bm{d_3}}^{\top}
\end{pmatrix*}  
-
\dfrac{1}{2}
\left(1 \ 1\right)
\begin{pmatrix*}[c]
{\bm{d_1}}^{\top} \\ 
{\bm{d_3}}^{\top}
\end{pmatrix*}  
\begin{pmatrix*}[c]
1 \\
1 \\
\end{pmatrix*}  
$,   
where the $i$th column of this equation represents the procedure of the Gram--Schmidt method to obtain the corresponding elements of $\bm{{b'}^d_i}$. Note that we denote the $l$th $(l = 1, \dots, n)$ row of $B'_d$, which parameterizes the $x$-coordinates of the J edges, as ${\bm{d^f_s}}^{\top}$ (resp. ${\bm{d^b_s}}^{\top}$) when $x_l$ is represented as $x_{h(s)+i}$ (resp. $x_{h(s)-i}$) in Eq.~(\ref{eq:parameterize_u_proposed}). Similarly, we denote the ($n+l$)th $(l = 1, \dots, n)$ row of $B'_d$, which parameterizes the $y$-coordinates of the J edges, as ${\bm{d^f_{n_v+s}}}^{\top}$ (resp. ${\bm{d^b_{n_v+s}}}^{\top}$) when $y_l$ is represented as $y_{h(s)+i}$ (resp. $y_{h(s)-i}$) in Eq.~(\ref{eq:parameterize_u_proposed}). 

After computing the necessary row vectors ${\bm{d^b_s}}^{\top}, {\bm{d^f_s}}^{\top} \ (s=1, 2, \dots, 2n_v)$, we can construct ${B'}_d$ as follows:
%\vspace{2mm}
\begin{equation}
%\footnotesize
\renewcommand{\arraystretch}{1.07}
\label{eq:B'd_IH51}
{B'}_d =
\begin{array}{rccll}
\ldelim[{19.5}{2pt}[] & & \cellcolor[gray]{.9}\bm{d_1}^{\top}           & \rdelim]{19.5}{2pt}[] &  :h(1)  \\
                    & & {\bm{d^f_1}}^{\top}                           &   & \rdelim\}{3}{10pt}[$h(1)+i \ (i = 1, \dots, k_1)$] \\ 
                    & & \scalebox{1.0}[0.8]{\vdots}                   &   & \\                  
                    & & {\bm{d^f_1}}^{\top}                           &   & \\
                    & & \cellcolor[gray]{.9}\bm{d_2}^{\top}           &   & :h(2)  \\ 
                    & & \frac{1}{2}(\bm{d_2} \! + \! \bm{d_3})^{\top} &   & \rdelim\}{3}{10pt}[$h(2)+i \ (i = 1, \dots, \lfloor \frac{k_2}{2} \rfloor)$] \\
                    & & \scalebox{1.0}[0.8]{\vdots}                   &   & \\                  
                    & & \frac{1}{2}(\bm{d_2} \! + \! \bm{d_3})^{\top} &   & \\
                    & & \frac{1}{2}(\bm{d_2} \! + \! \bm{d_3})^{\top} &   & :h(2) + \lfloor \frac{k_2+1}{2} \rfloor \\
                    & & \frac{1}{2}(\bm{d_2} \! + \! \bm{d_3})^{\top} &   & \rdelim\}{3}{10pt}[$h(3)-i \ (i = 1, \dots, \lfloor \frac{k_2}{2} \rfloor)$] \\
                    & & \scalebox{1.0}[0.8]{\vdots}                   &   & \\
                    & & \frac{1}{2}(\bm{d_2} \! + \! \bm{d_3})^{\top} &   & \\
                    & & \cellcolor[gray]{.9}\bm{d_3}^{\top}           &   & :h(3) \\ 
                    & & {\bm{d^f_3}}^{\top}                           &   & \rdelim\}{3}{10pt}[$h(3)+i \ (i = 1, \dots, k_1)$] \\ 
                    & & \scalebox{1.0}[0.8]{\vdots}                   &   & \\
                    & & {\bm{d^f_3}}^{\top}                           &   & \\
                    & & \cellcolor[gray]{.9}\bm{d_4}^{\top}           &   & :h(4) \\ 
                    & & \scalebox{1.0}[0.8]{\vdots}                   &   &
\end{array}
.
\end{equation} 
%\vspace{2mm}

In the final step, we need to orthogonalize the column vectors $\bm{{b'}^d_1}, \bm{{b'}^d_2}, \dots, \bm{{b'}^d_{m_d}}$ (the normalization is also performed here). Let $\bm{{b''}^d_1}, \bm{{b''}^d_2}, \dots, \bm{{b''}^d_{m_d}}$ be an orthonormal basis of $\mbox{span}(\bm{{b'}^d_1}, \bm{{b'}^d_2}, \dots, \bm{{b'}^d_{m_d}})$, which is obtained by the Gram--Schmidt method, and define the $2n \times m_d$ matrix ${B''}_d$ by
\begin{equation}
\label{eq:B''d}
{B''}_d = \left(\bm{{b''}^d_1} \ \bm{{b''}^d_2} \ \cdots \ \bm{{b''}^d_{m_d}} \right).
\end{equation}
Then, the matrix $B_{ik}$ is obtained as $({B''}_d \ B_s)$, and the tile shapes are finally parametrized as follows: 
\begin{equation}
\label{eq:u=B''d}
\bm{u} = {B''}_d \bm{\xi_d} + B_s \bm{\xi_s} = ({B''}_d \ B_s) 
\begin{pmatrix*}[c] 
\bm{\xi_d} \\
\bm{\xi_s} \\
\end{pmatrix*}
= B_{ik} \bm{\xi} 
.
\end{equation}

We now summarize the construction process of the matrix $B_{ik}$, as well as the time complexity of each step. When evaluating the time complexity, we approximate $m_d$ as $n_v$ ($m_d \leq 2 n_v$ for all isohedral types). Given an isohedral type $i \in I$, we only need to compute the matrix $B_v$ and then compute the row vectors ${\bm{d_{s}}}^{\top}, {\bm{d^b_s}}^{\top}, {\bm{d^f_s}}^{\top} \ (s=1, 2, \dots, 2n_v)$ as necessary, once at the beginning. These processes take $O({n_v}^3)$ and $O({n_v}^2)$ time, respectively, which are negligible compared to the time complexity of the overall processes. For each $k \in K_i$, matrix $B_{ik}$ is constructed by the following three steps.
\begin{enumerate}
\item 
Construct $B'_d$ simply by arranging the row vectors ${\bm{d_{s}}}^{\top}, {\bm{d^b_s}}^{\top}, {\bm{d^f_s}}^{\top} \ (s=1, 2, \dots, 2n_v)$, as shown in Eq.~(\ref{eq:B'd_IH51}). This process takes $O(n_v n)$ time, because $B'_d$ is a $2n \times m_d$ matrix.
\item 
Construct $B_s$ accordingly, as shown in Eq.~(\ref{eq:parameterize_u_proposed}). This process takes $O(n)$ time using an appropriate sparse matrix format, because the order of the nonzero elements of $B_s$ is $O(n)$. 
\item 
Orthonormalize the column vectors of $B'_d$ to generate ${B''}_d$. This process takes $O(n_v^2 n)$ time using the Gram--Schmidt method. 
\end{enumerate}
Considering that $n_v$ is constant, $B_{ik}$ is computed in $O(n)$ time. 

In fact, for any $i \in I$ and $k \in K_i$, we can construct $B_{ik}$ in $O(n)$ time in a similar way to that described above. Obviously, the matrix $B_v$ (Eq.~(\ref{eq:Bv})) can be obtained similarly for all isohedral types. As for Eq.~(\ref{eq:parameterize_u_proposed}), the tiling vertices are parameterized in the same way. The points on any S edge are parameterized in the same way, because each S edge can be deformed independently of the other tiling edges. On the contrary, each J edge has a corresponding J edge and there are several types of relationships between two corresponding J edges. The parameterization of the J edges depends on these relationship types. If a pair of J edges are glide-reflection symmetric with respect to the $x$-axis (or $y$-axis), these J edges are parameterized in the same way as Eq.~(\ref{eq:parameterize_u_proposed}). It is straightforward to parameterize a pair of J edges that are parallel to one another. However, it may be nontrivial to parameterize a pair of consecutive J edges that must form a specified angle. Similarly, it is easily understood that we can construct the matrix $B_{ik}$ according to the procedure described above for any isohedral type, unless the template does not include a pair of consecutive J edges that must form a specified angle. Therefore, we present another example for IH7 in Appendix A to describe how the matrix $B_{ik}$ is constructed in $O(n)$ time for an isohedral type including such J edges.

\subsection{Efficient calculation of the Procrustes distance} \label{sec:3_Procrustes}

\subsubsection{Projection method} \label{sec:3_Procrustes_Efficient}

Imahori et al. \cite{imahori2015escher} employed a projection method \cite{saad2011numerical}, which runs in $O(n^2)$ time, to compute the maximum eigenvalue of $B^{\top} V B$ and the corresponding eigenvector (Eq.~(\ref{eq:xiBVBxi})). We propose a more efficient calculation method that runs in $O(n)$ time. Although the details of the projection method used in \cite{imahori2015escher} were not presented in their paper, our method is regarded as a modified version of theirs (personal communication), which more effectively utilizes the feature of the matrix $B^{\top} V B$. 

The matrix $V$ defined in Eq.~(\ref{eq:V}) can be simplified as follows:
\begin{eqnarray}
\label{eq:V_simple}
V &=& \left(
    \renewcommand{\arraystretch}{1.1}	
    \begin{array}{cc}
      \bm{w_x}{\bm{w_x}}^{\top} & \ \ \bm{w_x}{\bm{w_y}}^{\top}  \\ 
      \bm{w_y}{\bm{w_x}}^{\top} & \ \ \bm{w_y}{\bm{w_y}}^{\top}  
    \end{array}
  \right)
+
    \left(
    \renewcommand{\arraystretch}{1.1}	
    \begin{array}{cc}
      \bm{w_y}{\bm{w_y}}^{\top}  & \ \ -\bm{w_y}{\bm{w_x}}^{\top}  \\ 
      -\bm{w_x}{\bm{w_y}}^{\top} & \ \  \bm{w_x}{\bm{w_x}}^{\top}  
    \end{array}
  \right) \nonumber \\
  &=& 
\begin{pmatrix*}[c]
\bm{w_x} \\ 
\bm{w_y}
\end{pmatrix*}  
\left({\bm{w_x}}^{\top} \ \ {\bm{w_y}}^{\top}\right)
+
\begin{pmatrix*}[c]
\bm{w_y} \\ 
-\bm{w_x}
\end{pmatrix*}  
\left({\bm{w_y}}^{\top} \ \ -{\bm{w_x}}^{\top}\right) \nonumber \\ 
&=& 
{\bm{w}} {\bm{w}}^{\top} + {\bm{w_c}} {\bm{w_c}}^{\top},
\end{eqnarray}
where 
$
\bm{w}=
\begin{pmatrix*}[c]
\bm{w_x} \\ 
\bm{w_y}
\end{pmatrix*}
$
and 
$
\bm{w_c}=
\begin{pmatrix*}[c]
\bm{w_y} \\ 
-\bm{w_x}
\end{pmatrix*}
$
.
Then, $B^{\top} V B$ can be also simplified as follows:
\begin{eqnarray}
\label{eq:BVB_simple}
B^{\top} V B &=& B^{\top} {\bm{w}} {\bm{w}}^{\top} B + B^{\top} {\bm{w_c}} {\bm{w_c}}^{\top} B \nonumber \\ 
&=& (B^{\top} {\bm{w}}){(B^{\top} {\bm{w}})}^{\top} + (B^{\top} {\bm{w_c}}){(B^{\top} {\bm{w_c}})}^{\top} \nonumber \\
&=& {\bm{p}} {\bm{p}}^{\top} + {\bm{p_c}} {\bm{p_c}}^{\top}, 
\end{eqnarray}
where 
$\bm{p}= B^{\top} {\bm{w}}$ 
and 
$\bm{p_c}= B^{\top} {\bm{w_c}}$. From the form of the above equation, $B^{\top} V B$ is a symmetric positive semidefinite matrix of maximum rank two. In general, when a matrix is given in the form of ${\bm{p}} {\bm{p}}^{\top} + {\bm{p_c}} {\bm{p_c}}^{\top}$, the eigenvalue problem can be reduced to the eigenvalue problem for a two-dimensional matrix in the subspace spanned by ${\bm{p}}$ and ${\bm{p_c}}$.

Let $B^{\top} V B$ be denoted as $C$ for simplicity, which is a square matrix of size $m$ ($B$ is a $2n \times m$ matrix). Then, the $m$-dimensional eigenvalue problem 
\begin{equation}
\label{eq:eigenvalue_problem_1}
C\bm{x} = \lambda \bm{x} 
\end{equation} 
can be reduced to a two-dimensional eigenvalue problem as follows. Let $\bm{e_1}$ and $\bm{e_2}$ be the orthonormal basis of $\mbox{span}(\bm{p}, \bm{p_c})$ and define the $m \times 2$ matrix $Q=(\bm{e_1} \ \bm{e_2})$. Then, any vector in $\mbox{span}(\bm{p}, \bm{p_c})$ is represented as $\bm{x}=Q\bm{y}$ with a two-dimensional parameter vector $\bm{y}$. Substituting this equation into Eq.~(\ref{eq:eigenvalue_problem_1}), we have $CQ\bm{y}=\lambda Q \bm{y}$. Then, multiplying by the matrix $Q^{\top}$ from the left, we have 
\begin{equation}
\label{eq:eigenvalue_problem_2}
Q^{\top}CQ\bm{y} = \lambda \bm{y},
\end{equation} 
where the relation $Q^{\top}Q = I$ (identity matrix) is used. Now, the eigenvalue problem in Eq.~(\ref{eq:eigenvalue_problem_1}) is reduced to a two-dimensional eigenvalue problem represented as Eq.~(\ref{eq:eigenvalue_problem_2}), because $Q^{\top}CQ$ is a square matrix of size two. The maximum eigenvalue $\lambda_{max}$ of Eq.~(\ref{eq:eigenvalue_problem_2}) and the corresponding eigenvector $\bm{y^*}$ are easily obtained. The maximum eigenvalue of Eq.~(\ref{eq:eigenvalue_problem_1}) is also $\lambda_{max}$ and the corresponding eigenvector $\bm{x^*}$ is obtained by $Q\bm{y^*}$.

We now evaluate the time complexity of the proposed calculation method. The calculation of $\bm{p}= B^{\top} {\bm{w}}$ is possible in $O(n_v n)$ time by using an appropriate sparse matrix format, because $B=({B''}_d, B_s)$ is a sparse matrix (only $B_s$ is sparse) whose order of nonzero elements is $O(n_v n)$. The same is true for $\bm{p_c}$. The matrix $Q$ is obtained in $O(n)$ time. To compute the matrix $Q^{\top}CQ$, we first transform this matrix as follows:
\begin{equation}
\label{eq:eigenvalue_problem_3}
Q^{\top}CQ = Q^{\top}B^{\top} V BQ =  Q^{\top}({\bm{p}} {\bm{p}}^{\top} + {\bm{p_c}} {\bm{p_c}}^{\top}) Q = (Q^{\top}\bm{p}){(Q^{\top}\bm{p})}^{\top} + (Q^{\top}\bm{p_c}){(Q^{\top}\bm{p_c})}^{\top}.
\end{equation}
The last matrix expression can be computed in $O(n)$ time, because both $Q^{\top}\bm{p}$ and $Q^{\top}\bm{p_c}$ are computed in $O(n)$ time. Consequently, it takes $O(n)$ time to compute the maximum eigenvalue of $B^{\top} V B$ and the corresponding eigenvector, because $n_v$ is a constant. 

\subsubsection{Use of the Euclidean distance} \label{sec:3_Procrustes_Euclid}

The Procrustes distance is scale- and rotation-invariant. The property of rotation-invariance is indispensable (see Section \ref{sec:2_similarity}) for certain isohedral types (IH2, IH3, IH5, and IH6), because the tile shapes $U$ can only appear in a specific orientation. On the contrary, this property is not necessary for other isohedral types (IH1, IH4, IH7, IH21, and IH28), because the tile shapes $U$ can be drawn on the $xy$-plane at any position, scale, and orientation for these isohedral types. In this case, the problem of finding the tile shape $U$ that minimizes $d^2(U,W)$ is equivalent to that of finding the tile shape $U$ that minimizes ${\left\| U-W \right\|}^2$. We refer to ${\left\| U-W \right\|}^2$ as the square of the Euclidean distance. When the Euclidean distance is applicable for an isohedral type, minimizing $d^2(U,W)$ under the constraint ${\bm u} = B {\bm \xi}$ is equivalent to the following optimization problem (except for the size):  
\begin{equation}
\label{eq:min_Bxi}
\mbox{minimize:} \ {\left\| B \bm{\xi} - \bm{w} \right\|}^2.   
\end{equation}
This is the least-squares problem, and the solution is given by 
\begin{equation}
\label{eq:xi*}
\bm{\xi^*} = {(B^{\top} B)}^{-1} B^{\top} \bm{w} = B^{\top} \bm{w}. 
\end{equation}
Then, the optimal tile shape $\bm{u^*}$ is obtained by $\bm{u^*} = B \bm{\xi^*} =  B B^{\top} \bm{w}$. The corresponding minimum value is calculated as follows:
\begin{eqnarray}
\label{eq:eval_Euclid}
{\left\| B \bm{\xi^*} - \bm{w} \right\|}^2 &=& 
{\bm{\xi^*}}^{\top} B^{\top} B \bm{\xi^*} - 2 {\bm{\xi^*}}^{\top} B^{\top} \bm{w} + {\bm{w}}^{\top} \bm{w} \nonumber \\
                    &=& {\bm{\xi^*}}^{\top} \bm{\xi^*} - 2 {\bm{\xi^*}}^{\top} \bm{\xi^*} + {\bm{w}}^{\top} \bm{w}   \nonumber \\
                    &=& -{\bm{\xi^*}}^{\top} \bm{\xi^*} + {\bm{w}}^{\top} \bm{w} 
.
\end{eqnarray}

When the Euclidean distance is applicable for an isohedral type, the procedure of computing $eval_{ikj}$ (Eq.~(\ref{eq:eval_ikj})) is replaced with Eq.~(\ref{eq:eval_Euclid}).
That is, 
\begin{equation}
\label{eq:eval_ikj_Euclid}
eval_{ikj} = \min_{\substack{{\bm u} = B_{ik} {\bm \xi} \\ {\bm \xi \in R^m} } } d^2(U,W_j) 
= \min_{\bm \xi \in R^m} {\left\| B_{ik} \bm{\xi} - \bm{w_j} \right\|}^2 
= -{\bm{\xi^*}}^{\top} \bm{\xi^*} + {\bm{w}}^{\top} \bm{w}, 
\end{equation}
where $\bm{\xi^*} = B_{ik}^{\top} \bm{w_j}$. According to our observations, computing the Euclidean distance is three to four times faster than computing the Procrustes distance.

\section{Efficient Exhaustive Search with Decomposition Relaxation} \label{sec:4}

Owing to the efficient calculation method for the matrix $B_{ik}$ (Section \ref{sec:3_B}), Algorithm \ref{alg:exhaustive_search} spends most of the execution time required for computing the maximum eigenvalue of ${B_{ik}}^{\top} V_j B_{ik}$ (or $B_{ik}^{\top} {\bm w_j}$ if the Euclidean distance is applicable) on computing $eval_{ikj}$, although it is also computed efficiently (Section \ref{sec:3_Procrustes}). To execute Algorithm \ref{alg:exhaustive_search} even more efficiently, we introduce a technique, called {\it decomposition relaxation}. This evaluates a lower bound on the value of $eval_{ikj}$ in the course of the search, in order to eliminate unnecessary computations of $eval_{ikj}$ that do not yield any improvement in $eval_{mim}$. 

Among the nine most general isohedral types, the order of $K_i$ reaches $O(n^3)$ for IH5 and IH6, and $O(n^4)$ for IH4, and Algorithm \ref{alg:exhaustive_search} spends most of the computation time on these isohedral types. Therefore, we only apply this technique to the isohedral types IH4, IH5, and IH6. Note that these isohedral types are the most important of the nine isohedral types, because they are highly flexible in expressing tile shapes, and the best tile shapes are usually obtained with one of these isohedral types. 

The basic ideas are common to isohedral types IH4, IH5, and IH6, but certain details are different. We first present the details of the decomposition relaxation technique for IH4 and describe how this idea is incorporated into Algorithm \ref{alg:exhaustive_search}. Then, we describe additional notes for IH5 and IH6. 

\subsection{Principle of decomposition relaxation and application to IH4} \label{sec:4_IH4}

From the template of IH4 shown in Fig. \ref{fig:IH_all}, all possible configurations for the assignment of $n$ points to the tiling edges are defined as $K_{4}=\{(k_1, k_2, k_3, k_4) \mid 0 \leq k_1, k_2, k_3, k_4, \ 2 k_1 + k_2 + k_3 + k_4 \leq n-6\}$, with $k_5$ determined by $k_5=n-6-(2 k_1 + k_2 + k_3 + k_4)$. In practice, we can apply the inequality constraint $k_4+k_5 < k_1+k_2$ to eliminate duplications arising from symmetry. The order of $K_4$ is $O(n^4)$ and we can use the Euclidean distance.

\begin{figure} [t] %%%%%%%%%%%%%%%%%%%%%%%%%%%%%
\centering
\includegraphics[scale=0.65,keepaspectratio,clip]{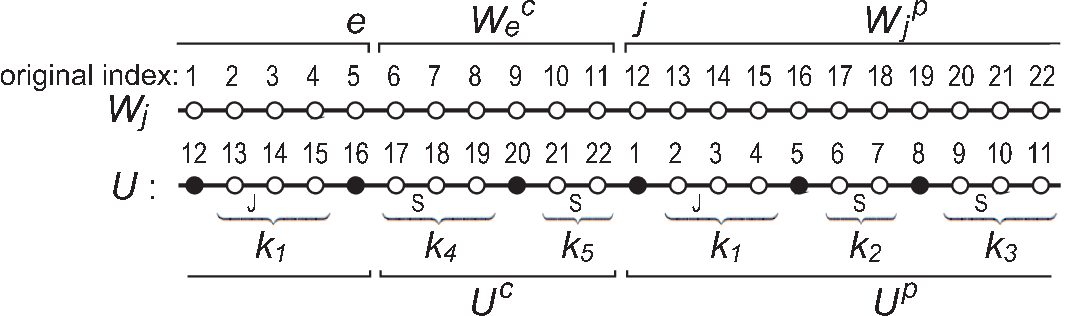}
\caption{Correspondence relationship between the tile polygon $U$ constrained by IH4 and the goal polygon $W_j \ (j=12)$.}
\label{fig:relation_UW}
\end{figure}

Let us denote $eval_{ikj}$ in Algorithm \ref{alg:exhaustive_search} as $eval(k_1, k_2, k_3, k_4; j)$, where the index for the isohedral type is omitted and $k$ is specifically represented. Eq.~(\ref{eq:eval_ikj_Euclid}) for IH4 is then rewritten as follows:
\begin{equation}
\label{eq:eval_full}
eval(k_1, k_2, k_3, k_4; j) 
= \min_{\substack{{\bm u} = B_{4k} {\bm \xi} \\ {\bm \xi \in R^m} } } d^2(U,W_j)
= \min_{\substack{{\bm u} = B_{4k} {\bm \xi} \\ {\bm \xi \in R^m} } } {\|U-W_j\|}^2.
\end{equation}
Recall that the $n$ points of the goal polygon $W$ are indexed clockwise by $1, 2, \dots, n$. We refer to this numbering scheme as the original index. Recall also that $W_j$ is the same polygon, but with the indices renumbered such that the $j$th point in the original index becomes the first point in the new index. Fig. \ref{fig:relation_UW} (see also the template of IH4 shown in Fig. \ref{fig:IH_all}) illustrates the correspondence relationship between the tile shape $U$ and the goal polygon $W_j \ (j=12)$, where the indices of the goal polygon $W_j$ are represented using the original index. For example, the first point of the tile shape $U$ corresponds to the $j$th point of the goal polygon $W$ when calculating Eq.~(\ref{eq:eval_ikj_Euclid}).  

The principle of the decomposition relaxation is simple. Let the tile shape $U$ be decomposed into two parts, $U^p$ and $U^c$, where the constraint conditions linking the two parts are relaxed. We then optimize each of the partial tile shapes independently, such that it is as close as possible to its corresponding partial goal shape under the relaxed constraints and compute the minimum value of the square of the Euclidean distance between the two shapes. Obviously, the sum of these values gives a lower bound on
$eval(k_1, k_2, k_3, k_4; j)$.

If the values for only $k_1$, $k_2$, and $k_3$ are given for IH4, we can parameterize partial tile shapes consisting of the first four tiling edges in the same way as in Eq.~(\ref{eq:u=Bxi}) (and, therefore, Eq.~(\ref{eq:u=B''d})). We denote this partial tile shape as $U^p$ and the corresponding matrix $B$ as $B^p_{4k}$, where a triplet $(k_1, k_2, k_3)$ is represented as $k$. In the same way, we denote by $W^p_j$ the corresponding partial goal shape, which starts from the $j$th point of $W$ (see Fig. \ref{fig:relation_UW}). Then, we define the following function: 
\begin{equation}
\label{eq:eval_partial}
eval_p(k_1, k_2, k_3; j) = \min_{\substack{{\bm u^p} = B^p_{4k} {\bm \xi} \\ {\bm \xi \in R^{m'}} } } d^2(U^p,W^p_j)
= \min_{\substack{{\bm u^p} = B^p_{4k} {\bm \xi} \\ {\bm \xi \in R^{m'}} } } {\|U^p-W^p_j\|}^2,
\end{equation}
where ${\bm{u_p}}$ is a vector representing the coordinates of $U^p$ and $m'$ is the dimension of the parameters. 

 %%%%%%%%%%%%%%%%%%%%%%%%%%%%%

Next, we consider the effect of the complementary part of the partial tile shape $U^p$ on $eval(k_1, k_2, k_3, k_4;j)$. We denote by $U^{c}$ the partial tile shape of the complementary part, which is constrained by the two consecutive S edges (the fifth and sixth tiling edges). We should note here that the ends of the two consecutive S edges are not included in $U^{c}$. In addition, we denote by $W^c_e$ the corresponding partial goal shape, which starts from the ($e+1$)th point of $W$, where the $e$th point of $W$ corresponds to the fifth tiling vertex of the tile shape $U$ (see Fig. \ref{fig:relation_UW}).

We see that the numbers of the points assigned to the two consecutive S edges of the partial tile shape $U^c$ are $k_4$ and $k_5$. Then, we can parameterize $U^c$ in the same way as in Eq.~(\ref{eq:u=B''d}). However, here, we need to parameterize the coordinates of the points on the two consecutive S edges, {\it not} including their ends, and $U^c$ is parameterized as follows. According to the construction process of the matrix $B_{ik}$ summarized at the end of Section \ref{sec:3_B}, (1--2) construct $B'_d$ and $B_s$ for the two consecutive S edges, including their ends, and then remove the four rows that parameterize the coordinates of the ends; (3) then, orthonormalize the column vectors of $B'_d$ to generate ${B''}_d$. We denote the constructed matrix $B$ as $B^{c}_{4k}$, where a pair $(k_4, k_5)$ is represented as $k$. Then, we define the following function: 
\begin{equation}
\label{eq:eval_c}
eval_{c}(k_4, k_5; e) 
= \min_{\substack{{\bm u^{c}} = B^{c}_{4k} {\bm \xi} \\ {\bm \xi \in R^{m''}} } } d^2(U^{c},W^{c}_e)
= \min_{\substack{{\bm u^{c}} = B^{c}_{4k} {\bm \xi} \\ {\bm \xi \in R^{m''}} } } {\|U^c-W^c_e\|}^2,
\end{equation}
where ${\bm{u_c}}$ is a vector representing the coordinates of $U^c$ and $m''$ is the dimension of the parameters. 

Given that ${\|U-W_j\|}^2 = {\|U^p-W^p_j\|}^2 + {\|U^c-W^c_e\|}^2$, from Eqs.~(\ref{eq:eval_partial}) and (\ref{eq:eval_c}), the following function gives a lower bound on $eval(k_1, k_2, k_3, k_4;j)$:
\begin{equation}
\label{eq:eval_lower}
eval_p(k_1, k_2, k_3; j) + eval_c(k_4,k_5; e),
\end{equation}
where $e=j+2k_1+k_2+k_3+5$ (subtract $n$ if $e > n$) and $k_5 = n-6-(2k_1+k_2+k_3+k_4)$. When we find that $eval_p(k_1, k_2, k_3; j) + eval_c(k_4,k_5; e) \geq eval_{min}$ in the course of the search of Algorithm \ref{alg:exhaustive_search}, we can omit the computation of $eval(k_1, k_2, k_3, k_4;j)$. To compute Eq.~(\ref{eq:eval_lower}) efficiently, we compute $eval_c(k_4,k_5; e)$ for $0 \leq k_4+k_5 \leq n-6$ and $1 \leq e \leq n$ at the beginning of the search. 

In Algorithm \ref{alg:BB}, we present the overall algorithm that incorporates the mechanism to eliminate unnecessary computations of $eval_{ikj}$ into the exhaustive search (Algorithm \ref{alg:exhaustive_search}). An important point to note is that for each combination of $(k_1, k_2, k_3, k_4) \in K_4$ and $j \in \{1, \dots, n\}$, we compute $eval(k_1, k_2, k_3, k_4;j)$ only when $eval_p(k_1, k_2, k_3; j) + eval_c(k_4,k_5;e) < eval_{min}$ (lines 7 and 9). In addition, we need to pay attention to line 8. For certain combinations of $(k_1, k_2, k_3, k_4) \in K_4$, there are cases in which it is unnecessary to compute $eval(k_1, k_2, k_3, k_4;j)$ for all $j = 1, 2, \dots, n$. In such cases, we do not need to construct the matrix $B_{4k}$, and we construct this matrix only when necessary.

\begin{algorithm}
\caption{: {\sc Efficient algorithm with decomposition relaxation (IH4)}}
\label{alg:BB}
\algsetup{
linenosize=\small,
linenodelimiter=:
}
\begin{algorithmic}[1]
\STATE $eval_{mim} := \infty$ (or inherited from the preceding search for other isohedral types);
\STATE Compute $eval_c(k_4, k_5; e)$ for all combinations of $k_4, k_5, e$;
\FOR{$0 \leq  2 k_1+k_2+k_3 \leq n-6$ (triple for-loop)}
\STATE Compute $eval_p(k_1,k_2,k_3; j) \rightarrow eval_j \ (1 \leq j \leq n)$
\FOR{$0 \leq k_4 \leq n-6-(2 k_1+k_2+k_3)$}
\FOR{$j = 1, 2, \dots, n$} 
\IF{$eval_j + eval_c(k_4,k_5;e) < eval_{min}$ (Eq.~(\ref{eq:eval_lower}))}
\STATE Construct $B_{4k}$ if not yet constructed;
\STATE Compute $eval(k_1,k_2,k_3,k_4; j) \rightarrow eval$
\IF{$eval < eval_{min}$}
\STATE $eval_{min} := eval$;
\STATE Compute $\bm{u^*} = B_{4k} {\bm \xi^*} \ ({\bm \xi^*}={B_{4k}}^{\top} {\bm w_j})$;
\ENDIF
\ENDIF
\ENDFOR
\ENDFOR
\ENDFOR

\end{algorithmic}
\end{algorithm}

\subsection{Application to IH5} \label{sec:4_IH5}

From the template of IH5 shown in Fig. \ref{fig:IH_all}, the order of $K_5$ is $O(n^3)$ and the Procrustes distance must be used. The same approach of calculating the lower bound on $eval_{ikj}$ as that applied to IH4 can be also used for IH5, because (i) if the values for only $k_1$ and $k_2$ are given, we can parameterize partial tile shapes consisting of the first four tiling edges, and (ii) the complementary part (consisting of two consecutive S edges) can be also parametrized in the same way as for IH4. We denote $eval_{ikj}$ as $eval(k_1, k_2, k_3; j)$ in the IH5 case and a lower bound of $eval(k_1, k_2, k_3; j)$ is then given by 
\begin{equation}
\label{eq:eval_lower_IH5}
eval_p(k_1, k_2; j) + eval_c(k_3,k_4; e),
\end{equation}
where $e=j+2k_1+2k_2+5$ (subtract $n$ if $e > n$) and $k_4 = n-6-(2k_1+2k_2+k_3)$. 

We must redefine the Procrustes distance, because in the original Procrustes distance (Eq.~(\ref{eq:procrustes})), the size of the goal shape $W$ is normalized by dividing $W$ by $\|W\|$. However, the normalization of the image size is problematic, because different scaling factors are applied to the complete goal shape $W$ and the partial goal shapes $W^p_j$ and $W^c_e$. Therefore, we redefine the Procrustes distance as follows:
\begin{equation}
\label{eq:procrustes_no_scale}
d^2(U,W) = \min_{\theta} { \left\| R(\theta)U - W  \right\| }^2.
\end{equation}
By the same calculation as that used to transform from Eq.~(\ref{eq:procrustes}) to Eq.~(\ref{eq:V}), we have 
\begin{eqnarray}
\label{eq:procrustes_redefine}
d^2(U,W) &=& {\bm w}^\top {\bm w} + {\bm u}^\top {\bm u} - 2 \sqrt{{\bm u}^\top V {\bm u }} \nonumber \\
         &=& {\bm w}^\top {\bm w} + {\bm \xi}^\top {\bm \xi} - 2 \sqrt{{\bm \xi}^{\top} B^{\top} V B {\bm \xi}}.
\end{eqnarray}
By solving the equation $\dfrac{\partial d^2(U,W)}{\partial \bm \xi} = {\bm 0}$, we finally have the minimum value $-\lambda + {\bm w}^{\top} {\bm w}$ at ${\bm \xi} = {\bm \xi^*}$, where $\lambda$ is the maximum eigenvalue of $B^{\top}VB$ and ${\bm \xi^*}$ is the corresponding eigenvector, whose length is $\sqrt{\lambda}$. As a matter of course, the vector ${\bm \xi^*}$ obtained here is the same as that derived from the original Procrustes distance, except that the length of ${\bm \xi^*}$ is determined here. With this change, we need to modify Eq.~(\ref{eq:d2_procrustes}) as follows:
\begin{equation}
\label{eq:d2_procrustes_modi}
\min_{\substack{{\bm u} = B {\bm \xi} \\ {\bm \xi \in R^m} } } d^2(U,W) = {\|W\|}^2 - \lambda.
\end{equation}

\subsection{Application to IH6} \label{sec:4_IH6}

From the template of IH6 shown in Fig. \ref{fig:IH_all}, the order of $K_6$ is $O(n^3)$ and the Procrustes distance must be used. We denote $eval_{ikj}$ as $eval(k_1, k_2, k_3; j)$ in the IH6 case. A similar approach of calculating the lower bound on $eval_{ikj}$ to that used for IH4 is applicable, but certain aspects need to be modified. 

\begin{figure} [t] %%%%%%%%%%%%%%%%%%%%%%%%%%%%%
\centering
\includegraphics[scale=0.65,keepaspectratio,clip]{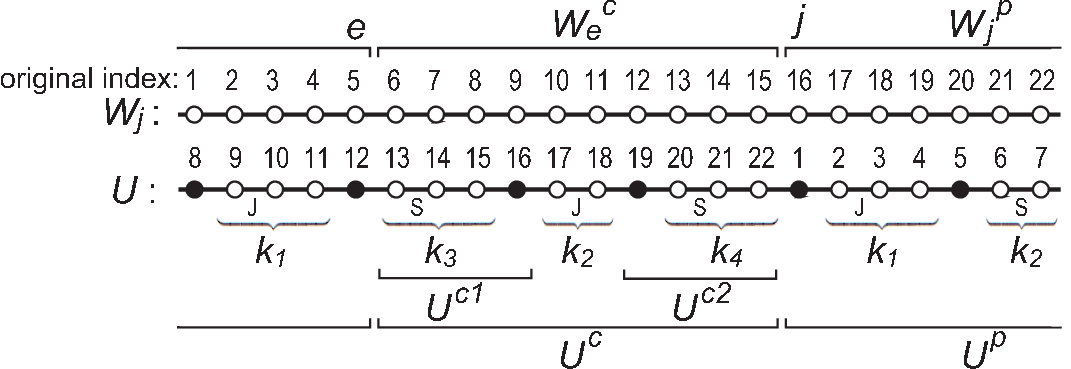}
\caption{Correspondence relationship between the tile polygon $U$ constrained by IH6 and the goal polygon $W_j \ (j=16)$.}
\label{fig:relation_UW_IH6}
\end{figure}

Fig. \ref{fig:relation_UW_IH6} (see also the template of IH6 shown in Fig. \ref{fig:IH_all}) illustrates the correspondence relationship between $U$ and $W_j \ (j=16)$, where the indices of the goal polygon are represented using the original index. If the values for only $k_1$ and $k_2$ are determined, we can parameterize the partial tile shape $U^p$ consisting of only the first three tiling edges (although $k_2$ specifies the number of points assigned to the fifth tiling edge), because the number of points assigned to the fourth tiling edge $k_3$ is unknown. This problem occurs, regardless of how we select the first tiling edge to define the template. We can compute $eval_p(k_1, k_2; j)$ in a similar way to that in Eq.~(\ref{eq:eval_partial}), where the redefined Procrustes distance must be used.

The complementary part $U^c$ consists of the fourth to sixth tiling edges (not including their ends), which are parameterized with the three parameters $k_3$, $k_2$, and $k_4$. Although we can compute $eval_c(k_3, k_2, k_4; e)$ in a similar way to that in Eq.~(\ref{eq:eval_c}), computing this value for all possible argument values seems to require almost the same computational effort as that of $eval(k_1, k_2, k_3; j)$. In fact, we can compute $eval_c(k_3, k_2, k_4; e)$ more efficiently as follows. The complementary part $U^c$ is constrained by the three consecutive edges arranged in the order of S, J, and S edges. Let us further define two partial tile shapes, $U^{c1}$ and $U^{c2}$, as illustrated in Fig. \ref{fig:relation_UW_IH6}. We define $eval_{c1}(k_3; e)$ as the minimum value of the square of the Procrustes distance between $U^{c1}$ and the corresponding partial goal shape when $U^{c1}$ is constrained only by the S edge (but one end is not included), and the start point of this S edge corresponds to the $e$th point of the goal polygon $W$. We define $eval_{c2}(k_4; j)$ similarly. 

We can see that the J edge in $U^c$ does not have a corresponding J edge in the same partial tile shape, and therefore this J edge alone does not impose any constraint on the points of $U^c$ (this is also true for $U^p$). Therefore, the points of $U^{c1}$ ($U^{c2}$) are constrained only by the corresponding S edge and the remaining parts (the points on the J edge, not including its ends) completely overlap the corresponding points of the goal polygon after $U^c$ is shaped as closely as possible to the partial goal polygon $W^c_e$ under the imposed constraints. Consequently, $eval_c(k_3, k_2, k_4; e)$ is calculated simply as the sum of $eval_{c1}(k_3; e)$ and $eval_{c2}(k_4; j)$. Finally, a lower bound of $eval(k_1, k_2, k_3; j)$ is obtained by 
\begin{equation}
\label{eq:eval_lower_IH6}
eval_p(k_1, k_2; j) + eval_{c1}(k_3; e) + eval_{c2}(k_4; j),
\end{equation}
where $e=j+2k_1+k_2+4$ (subtract $n$ if $e > n$) and $k_4 = n-6-(2k_1+2k_2+k_3)$.

 %%%%%%%%%%%%%%%%%%%%%%%%%%%%%

\section{Experimental  Results} \label{sec:5}

We present our experimental results of evaluating the proposed algorithm for the exhaustive version of Koizumi and Sugihara's formulation of the Escherization problem in terms of the computation time and the quality of the obtained tile shapes.

\subsection{Experimental settings} \label{sec:5_setting}

We implemented the proposed algorithm in C++ with the Eigen library (template headers for linear algebra, matrix, and vector operations) in Ubuntu 14.04 Linux. We executed the program on an PC with an Intel Core i7-4790 3.60~GHz CPU.

We applied the proposed algorithm with several different settings to analyze the impact of each of three enhancements (see also Section \ref{sec:3_Exhaustive}): (1) an efficient calculation method of the matrix $B_{ik}$, (2) an efficient calculation method of the Procrustes distance, and (3) a technique to skip unnecessary computations of $eval_{ikj}$. Specifically, we compared the following three algorithms. 
\begin{description}
\item[Algorithm A:]
Algorithm \ref{alg:exhaustive_search} with enhancement (2).
\item[Algorithm B:]
Algorithm \ref{alg:exhaustive_search} with enhancements (1) and (2).
\item[Algorithm C:]
Algorithm \ref{alg:BB} with enhancements (1), (2), and (3).
\end{description}
We should note that we did not implement Algorithm \ref{alg:exhaustive_search} with no enhancements, and regard the results obtained by Algorithm A as a baseline, because we did not know the details of how the Procrustes distance was computed in previous studies \cite{imahori2012local,imahori2015escher}.

We applied Algorithms A, B, and C to two goal polygons, ``cat'' and ``hippocampus'', with 60 and 120 points located at equal intervals on the boundaries. The left side of Fig. \ref{fig:shape_cat_hipp} shows the two goal polygons with 60 points, from which the two goal polygons with 120 points were generated by placing an additional point between each pair of adjacent points. We used the redefined Procrustes distance (Eq.~(\ref{eq:procrustes_no_scale})) instead of the original one, because this enables comparing the distances calculated by the Euclidean distance and the Procrustes distance on the same scale. 

\subsection{Results} \label{sec:5_result}

Table \ref{table:feature} lists the order of $K_i$ and the distance measure used for each isohedral type. The table also lists the square of the distance between the optimal tile shape $U$ and the goal shape $W$ for each isohedral type. 

Table \ref{table:computation_time} lists the execution times (in seconds) of the three algorithms on the goal polygon ``cat'' with 60 and 120 points. Each row labeled IH$i$ gives the execution time when using only the isohedral type IH$i$. The row labeled ``All'' gives the total execution time when the nine most general isohedral types are applied in the order listed in the table. For Algorithms A and B, the total execution times are almost equal to the sum of the execution times of the individual isohedral types. For Algorithm C, the total execution time is shorter than the sum of the individual execution times, because the value of $eval_{min}$ (in Algorithm \ref{alg:BB}) is inherited from the preceding search for the other isohedral types, and unnecessary computations of $eval_{ikj}$ are further eliminated. The execution times on the goal polygon ``hippocampus'' were almost the same as those on the goal polygon ``cat'' and these results are omitted. 

\begin{table*}  [t] %%%%%%%%%%%%%%%%%%%%%%%%%%
\begin{center}
\caption{Order of $K_i$, the distance metric used, and the best evaluation value for each isohedral type.}
\label{table:feature}
\begin{tabular}{r|rr|rr|rr}
\hline	
      & Order      & Distance  &   \mul{4}{c}{The square of the minimum distance} \\ 
\cline{4-7}	
IH$i$ & of $K_i$   & metric    & cat(60) &  cat(120) & hipp.(60) & hipp.(120)  \\
\hline	
IH4   &  $O(n^4)$  & Eucl. &  10187.3 &  16469.1   &  21868.6 &  39974.7  \\
IH5   &  $O(n^3)$  & Proc. &  17792.7 &  32594.3   &  24312.3 &  41886.8  \\
IH6   &  $O(n^3)$  & Proc. &  22403.1 &  39499.1   &  18994.2 &  32989.6  \\
IH1   &  $O(n^2)$  & Eucl. & 115178.5 & 222848.9   &  79547.7 & 108226.8  \\
IH2   &  $O(n^2)$  & Proc. & 124939.2 & 223379.9   &  51370.7 &  76777.6  \\
IH3   &  $O(n^2)$  & Proc. &  45009.1 &  88606.9   &  25259.6 &  41507.2  \\
IH7   &  $O(n^2)$  & Eucl. &  63901.0 & 125382.9   & 145286.6 & 297239.6  \\
IH21  &  $O(n^2)$  & Eucl. &  85008.5 & 162412.1   &  39140.7 &  52378.1  \\
IH28  &  $O(n^2)$  & Eucl. &  83402.6 & 149559.4   &  79555.3 & 125427.4  \\
\hline	
\end{tabular}
\end{center}
\end{table*}

\begin{table*}  [t] %%%%%%%%%%%%%%%%%%%%%%%%%%
\begin{center}
\caption{Execution time (s) of the three algorithms.}
\label{table:computation_time}
\begin{tabular}{r|rrr|rrr}
\hline	
      &  \mul{3}{c|}{cat(60)}    &  \mul{3}{c}{cat(120)}         \\
\cline{2-7}	
IH$i$ &  Alg. A &Alg. B & Alg. C &  Alg. A     &Alg. B   & Alg. C \\
\hline	
IH4   &   88.76 &  3.65 &  0.20  &   11187.72  & 194.22  &  3.99  \\
IH5   &    5.44 &  0.70 &  0.10  &     333.07  &  18.83  &  1.28  \\
IH6   &    5.44 &  0.70 &  0.13  &     333.82  &  18.83  &  2.54  \\
IH1   &    0.24 &  0.02 &  =B    &       7.26  &   0.18  &  =B    \\
IH2   &    0.29 &  0.04 &  =B    &       8.37  &   0.53  &  =B    \\
IH3   &    0.30 &  0.04 &  =B    &       8.53  &   0.53  &  =B    \\
IH7   &    0.49 &  0.03 &  =B    &      14.42  &   0.40  &  =B    \\
IH21  &    0.48 &  0.03 &  =B    &      14.26  &   0.39  &  =B    \\
IH28  &    0.49 &  0.03 &  =B    &      14.22  &   0.39  &  =B    \\
\hline
All   &  102.10 &  5.13 &  0.55  &   11875.16  & 235.44  &  9.01  \\
\hline
\end{tabular}
\end{center}
\end{table*}

\begin{figure} [t] %%%%%%%%%%%%%%%%%%%%%%%%%%%%%
\centering
\includegraphics[scale=0.65,keepaspectratio,clip]{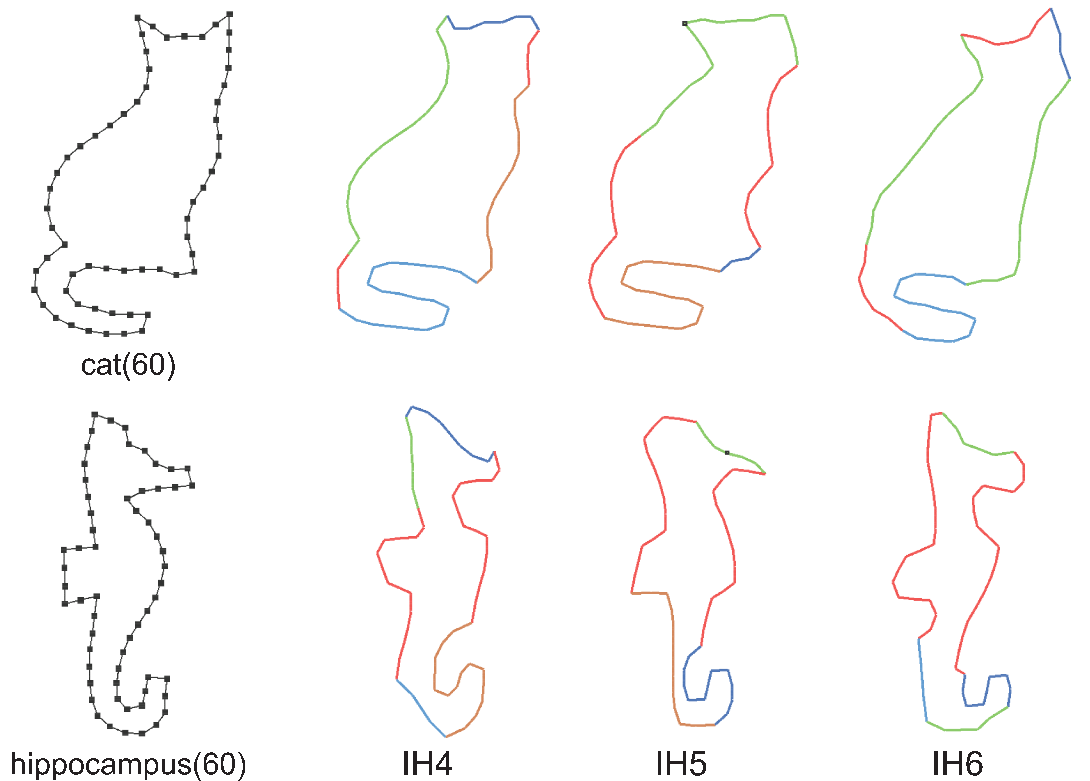}
\caption{Goal polygons ``cat'' and ``hippocampus'' ($n=60$) and the optimal tile shapes obtained with IH4, IH5, and IH6.}
\label{fig:shape_cat_hipp}
\end{figure}

\begin{figure} [t] %%%%%%%%%%%%%%%%%%%%%%%%%%%%%
\centering
\includegraphics[scale=0.65,keepaspectratio,clip]{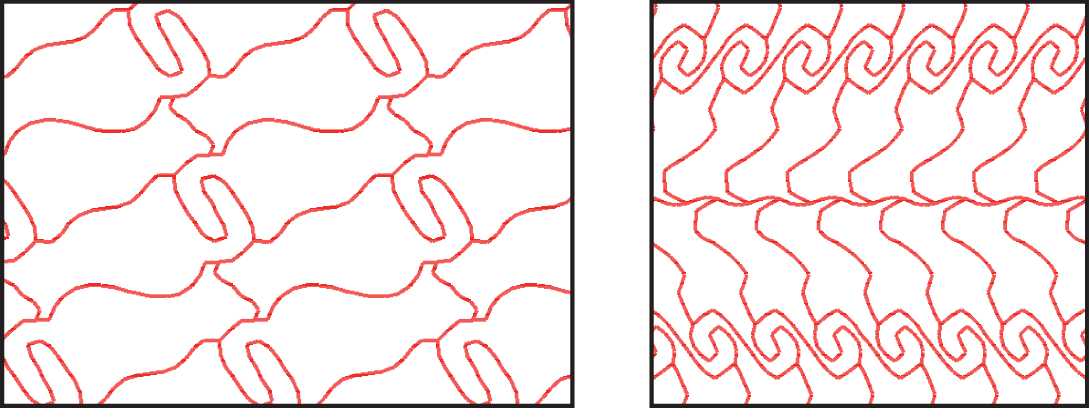}
\caption{Tilings generated from the tile shapes for cat(60) (IH4) and for hippocampus(60) (IH5).}
\label{fig:tiling_cat_hipp}
\end{figure}

Fig. \ref{fig:shape_cat_hipp} shows the optimal tile shapes obtained with the isohedral types IH4, IH5, and IH6, on the two goal polygons with 60 points. The optimal tile shapes for the goal polygons with 120 points are omitted, because nearly the same tile shapes were obtained. The tile shapes are drawn in different colors for each tiling edge, where the colors correspond to those in the templates shown in Fig. \ref{fig:IH_all}. The optimal tile shapes occasionally have an intersection, because the constraint conditions expressed as Eq.~(\ref{eq:Au=0}) do not exclude this. Therefore, for each isohedral type, we selected the best tile shape that had no intersection (the results on hippocampus(60) with IH6 apply to this situation). In Fig. \ref{fig:tiling_cat_hipp}, we present two tilings generated from the tile shapes obtained with IH4 for cat(60) and with IH5 for hippocampus(60). 

From Table \ref{table:feature}, we can see that good tile shapes with small distances tend to be obtained by the isohedral types IH4, IH5, and IH6 (IH3 is also suitable for hippocampus(60) and hippocampus(120)). Of course, other isohedral types may be best-suited for other goal polygons, but from our observations, the minimum distances were often obtained by either IH5, IH6, or especially IH4, for most of the goal polygons we tested. The reason for this is that the degree of freedom of the parameterized tile shapes increases with the order of $K_i$. 

From Table \ref{table:computation_time}, we can see that the execution time of IH4 is particularly large, and those of IH5 and IH6 are the next largest. This is also due to the difference in the order of $K_i$. When $n=60$, the total execution time of Algorithm A is 102 s, most of which is spent on IH4. We should note here that Algorithm A includes enhancement (2) and it takes more execution time if the efficient calculation method of the Procrustes distance is not used. For example, in \cite{imahori2015escher}, it took approximately 0.8 s to execute the Koizumi and Sugihara's original algorithm on a goal polygon with 60 points, whereas it took approximately 0.005 s to execute Algorithm A on cat(60) when only the same number of points were assigned to every tiling edge. By introducing enhancement (1), i.e., Algorithm B, the total execution time was reduced to 5.13 s, which is approximately 20 times faster than that of Algorithm A. Moreover, by introducing enhancement (3), i.e., Algorithm C, the total execution time was further reduced to only 0.55 s, which is approximately 9 times and 180 times faster than those of Algorithms B and A, respectively. When $n=120$, the acceleration effect becomes more prominent. In this case, the total execution time of Algorithm A was 11875.16 s, which is not a tolerable waiting time for a software application. The execution time of Algorithm C was reduced to 9.01 s, which is approximately 1300 times faster than that of Algorithms A. In our observations, the execution time also depended on the complexity of the goal polygons, where it was at most 1.5 times greater than the execution time listed in Table \ref{table:computation_time} for highly complex goal polygons $(n=60)$. 

Additionally, we applied the proposed algorithm to 24 goal polygons with 58--126 points, and we present the tile shapes and tilings obtained in the supplementary material.

 %%%%%%%%%%%%%%%%%%%%%%%%%%%%%

 %%%%%%%%%%%%%%%%%%%%%%%%%%%%%

\subsection{Discussion} \label{sec:5_discussion}

\begin{figure} [t] %%%%%%%%%%%%%%%%%%%%%%%%%%%%%
\centering
\includegraphics[scale=0.33,keepaspectratio,clip]{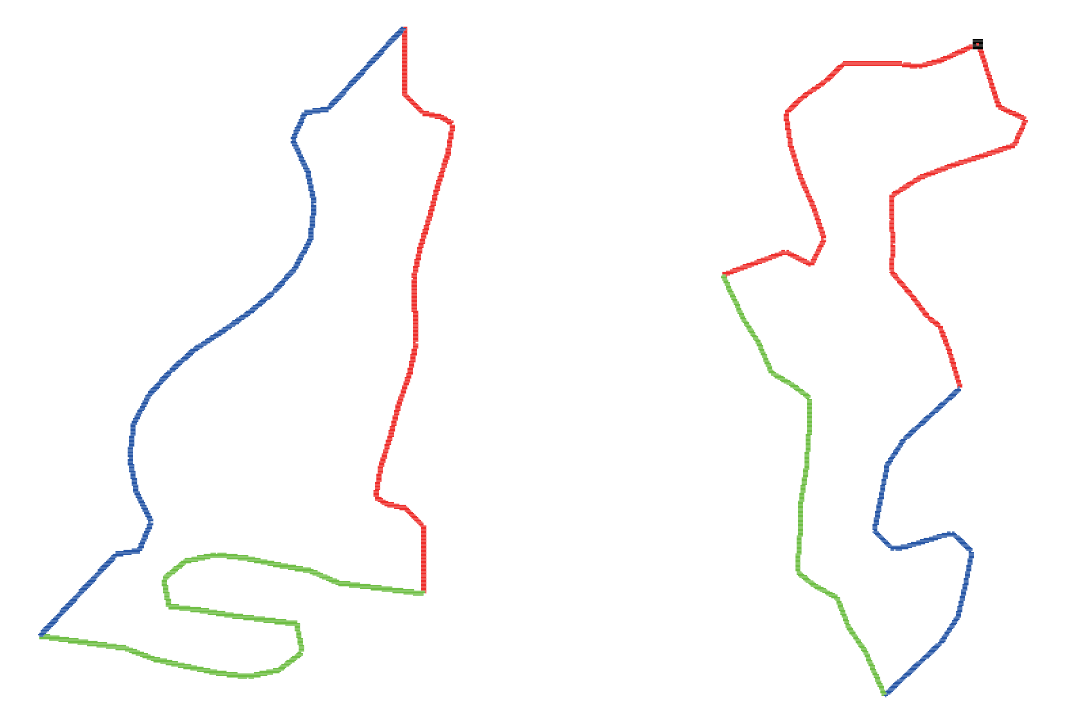}
\caption{Tile shapes obtained by  Koizumi and Sugihara's original formulation for the goal polygons cat(60) and hippocampus(60).}
\label{fig:shape_koizumi}
\end{figure}

We now discuss the quality of the tile shapes obtained by the exhaustive version of Koizumi and Sugihara's formulation. Fig. \ref{fig:shape_koizumi} shows the best tile shapes for the goal polygons cat(60) and hippocampus(60) obtained by Koizumi and Sugihara's original formulation. We can see that the quality of these tile shapes is poor compared to that of the tile shapes obtained by the exhaustive formulation (see Fig. \ref{fig:shape_cat_hipp}). As stated in Section \ref{sec:3_Motivation}, the reason for this is obvious: we need to search for a proper arrangement of points for the goal polygon to obtain good results, as long as the same number of points are assigned to every tiling edge.

\begin{figure} [t] %%%%%%%%%%%%%%%%%%%%%%%%%%%%%
\centering
\includegraphics[scale=0.65,keepaspectratio,clip]{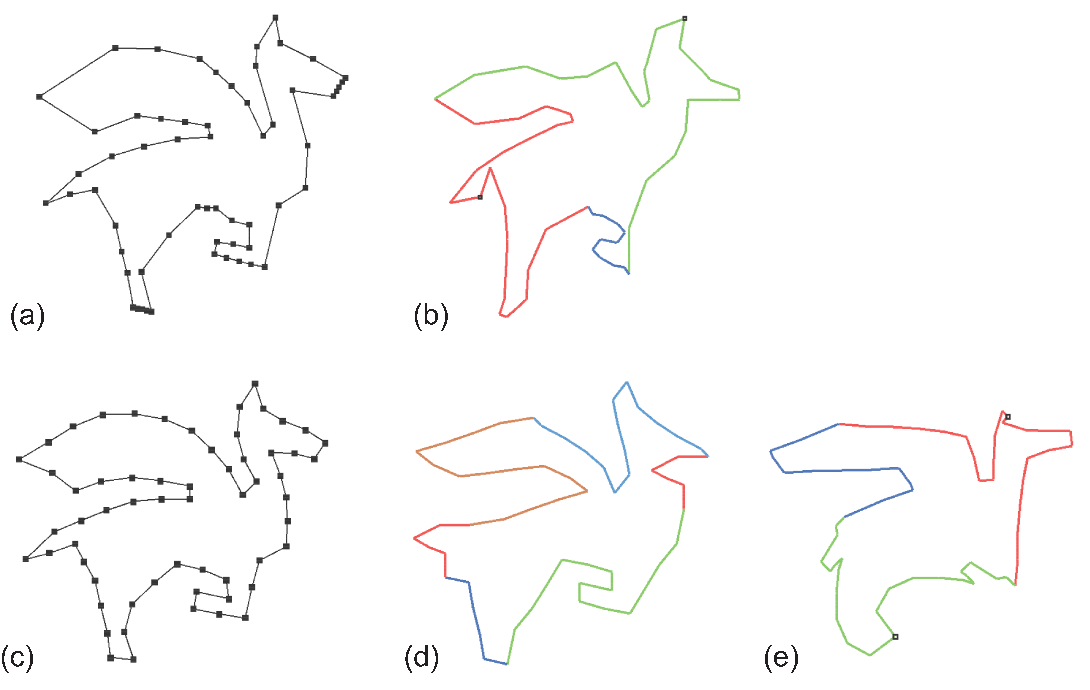}
\caption{Comparison with Imahori and Sakai's method: (a,b) Imahori and Sakai's method, (c,d) the proposed method, and (c,e) Koizumi and Sugihara's original method.}
\label{fig:shape_pegasus}
\end{figure}

To address this problem, Imahori and Sakai \cite{imahori2012local} proposed a local search-based method to find a good set of points for the goal polygon (see Section \ref{sec:3_Motivation}). Fig. \ref{fig:shape_pegasus} shows comparison results of Imahori and Sakai's method and the proposed method for a goal figure ``Pegasus''. Fig. \ref{fig:shape_pegasus}(a) and (b) show the goal polygon ($n=60$) generated by Imahori and Sakai's method and the tile shape obtained by Koizumi and Sugihara's original formulation with this goal polygon\footnote{We created this goal polygon by measuring the coordinates of the points from the figure in their paper, and the obtained tile shape is slightly different from theirs.}. Fig. \ref{fig:shape_pegasus}(c) and (d) show the goal polygon ($n=60$), whose points are arranged at nearly equal intervals and the tile shape obtained by the exhaustive Koizumi and Sugihara's formulation with this goal polygon. In addition, Fig. \ref{fig:shape_pegasus}(e) shows the tile shape obtained by Koizumi and Sugihara's original formulation with this goal polygon. Imahori and Sakai's method seems to perform well in overcoming the weakness of Koizumi and Sugihara's method, but it is unknown whether good results can be obtained stably, because this method is based on local search (only the best result was presented in the literature). According to the literature, the execution time of Imahori and Sakai's method was a few hundred seconds (on a PC with Pentium D@3.40~GHz CPU). Our efficient calculation methods of the matrix $B$ and the Procrustes distance could also be useful to accelerate the local search procedure of Imahori and Sakai's method. 

 %%%%%%%%%%%%%%%%%%%%%%%%%%%%%

 %%%%%%%%%%%%%%%%%%%%%%%%%%%%%

\begin{figure} [t] %%%%%%%%%%%%%%%%%%%%%%%%%%%%%
\centering
\includegraphics[scale=0.60,keepaspectratio,clip]{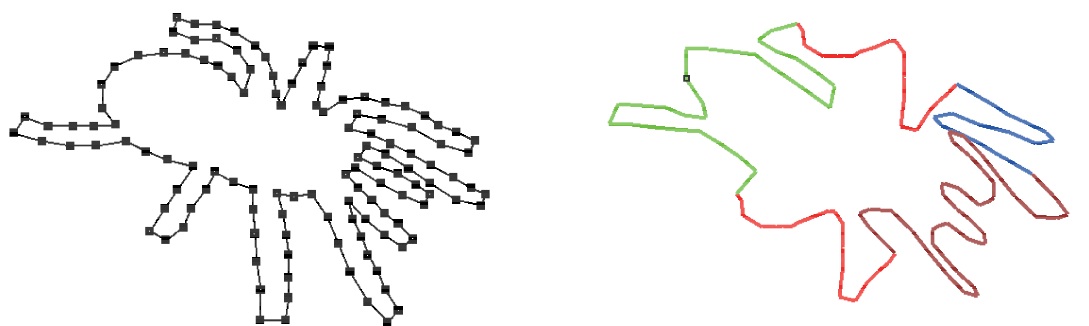}
\caption{Goal polygons ``spider'' ($n=126$) and the optimal tile shape.}
\label{fig:shape_spider}
\end{figure}

We also show a case in which a satisfactory tile shape cannot be obtained by the proposed method. Fig. \ref{fig:shape_spider} shows a goal polygon ``spider'' $(n=126)$, which is a challenging goal polygon, and the optimal tile shape obtained. The optimal tile shape belongs to IH5 and the execution time was 14.8 s. As we can see from the figure, the optimal tile shape is unsatisfactory. This means that there exists no figure that can tile the plane and is sufficiently similar to the goal polygon. Unfortunately, in such cases, the goal polygon itself need to be modified significantly.

 %%%%%%%%%%%%%%%%%%%%%%%%%%%%%

We further discuss the expected directions of related research in the future. Ono et al. \cite{ono2015figure} developed an interactive genetic algorithm (GA) that uses Koizumi and Sugihara's original algorithm as a subroutine. This system supports users in creating Escher-like tilings for an input goal figure. The basic idea is to evolve good sets of points for the goal polygon using the GA, which are evaluated by the user according to the resulting tile shapes obtained by Koizumi and Sugihara's algorithm. By repeating this operation, the system finally outputs tile shapes that reflect the user's preferences. As in their work, some type of trial-and-error process would be indispensable for finding satisfactory Escher-like tilings. Our method also contributes to this research direction \cite{ono2014creation,liu2017artwork}, because it enables finding the optimal tile shape from a huge search space within a short computation time.

\section{Conclusions} \label{sec:6}
In Koizumi and Sugihara's original formulation of the Escherization problem, the same number of points were assigned to every tiling edge, which led to the failure to obtain good tile shapes unless the points were properly arranged on the boundary of the goal figure. To solve this problem, as well as to obtain better-quality tile shapes, we have considered all possible combinations of the assignment of points to the tiling edges for the nine most general isohedral types, in an exhaustive version of Koizumi and Sugihara's formulation. To deal with an enormous increase in computation time, we have proposed efficient calculation methods for the matrix $B$ and the Procrustes distance, which enable computing the optimal tile shape in $O(n^2)$ time for each of the assignments of the $n$ points to the tiling edges, as opposed to $O(n^3)$ time in the original method. Moreover, we have proposed an efficient algorithm to eliminate unnecessary computations of the Procrustes distance by evaluating its lower bound during the search. The proposed algorithm has succeeded in reducing the execution time by factors of at least 180 and 1300 with respect to the original method for $n = 60$ and $120$, respectively. Consequently, our algorithm has found optimal tile shapes in 0.55 s ($n=60$) and 9.01 s ($n=120$).

\section*{Acknowledgment}
We would like to thank S. Sakai and S. Kawade for the goal figures used in the experiments. This work was supported by JSPS KAKENHI Grant Number 17K00342.

%\bibliographystyle{abbrv}
%BibTeX users please use one of
\bibliographystyle{spmpsci}      % mathematics and physical sciences
\bibliography{arxiv}

\bigskip

\section*{Appendix A: Parameterization of tile shapes for IH7}

For any $i \in I$ and $k \in K_i$, the tile shape $U$ can be parameterized and the matrix $B_{ik}$ can be constructed in $O(n)$ time, as described in Section \ref{sec:3_B}. We present another example of how the tile shape $U$ is parameterized for the isohedral type IH7, which we consider the most difficult to understand. 

\begin{figure} [t] %%%%%%%%%%%%%%%%%%%%%%%%%%%%%
\centering
\includegraphics[scale=0.60,keepaspectratio,clip]{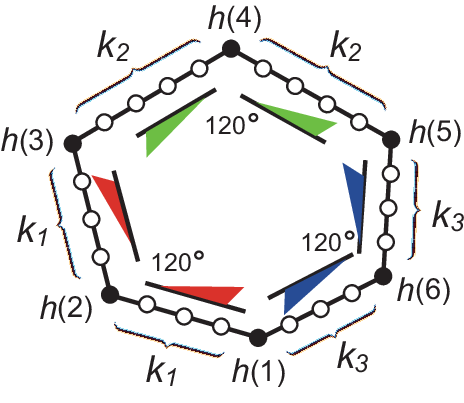}
\caption{Template of IH7 for a specific assignment of the points to the tiling edges.}
\label{fig:template_point_IH7}
\end{figure}

Fig. \ref{fig:template_point_IH7} shows the template of IH7 expressed in a similar way to that in Fig. \ref{fig:template_point}. The constraint conditions imposed on the tiling vertices are expressed by 
\begin{equation}
\label{eq:constraint_vertex_IH7}
\left\{
\begin{array}{lll}
x^v_{3} - x^v_{2} & = & \cos \theta (x^v_{1} - x^v_{2}) - \sin \theta (y^v_{1} - y^v_{2}) \\ 
y^v_{3} - y^v_{2} & = & \sin \theta (x^v_{1} - x^v_{2}) + \cos \theta (y^v_{1} - y^v_{2}) \\ 
x^v_{5} - x^v_{4} & = & \cos \theta (x^v_{3} - x^v_{4}) - \sin \theta (y^v_{3} - y^v_{4}) \\ 
y^v_{5} - y^v_{4} & = & \sin \theta (x^v_{3} - x^v_{4}) + \cos \theta (y^v_{3} - y^v_{4}) \\ 
x^v_{1} - x^v_{6} & = & \cos \theta (x^v_{5} - x^v_{6}) - \sin \theta (y^v_{5} - y^v_{6}) \\ 
y^v_{1} - y^v_{6} & = & \sin \theta (x^v_{5} - x^v_{6}) + \cos \theta (y^v_{5} - y^v_{6}) 
\end{array}%
\right.
,
\end{equation}
where $\theta = 120^{\circ}$. Then, we can obtain the matrix $B_v$ in the same way as in Eq.~(\ref{eq:Bv}). 

The $xy$-coordinates of the $n$ points are constrained by the following equations (only the constraints for the first and second tiling edges are shown):
\begin{equation}
\label{eq:constraint_edge_IH7}
\left\{
\begin{array}{lll}
x_{h(2)+i}-x_{h(2)} &=& \cos \theta (x_{h(2)-i}-x_{h(2)}) - \sin \theta (y_{h(2)-i}-y_{h(2)}) \ (i = 1, \dots, k_1)\\
y_{h(2)+i}-y_{h(2)} &=& \sin \theta (x_{h(2)-i}-x_{h(2)}) + \cos \theta (y_{h(2)-i}-y_{h(2)}) \ (i = 1, \dots, k_1)\\
& \vdots &
\end{array}%
\right.
.
\end{equation}

 %%%%%%%%%%%%%%%%%%%%%%%%%%%%%
%\colorbox[gray]{.9}{$\color[gray]{0.9}{{}^1 }$} \\
By parameterizing $x_{h(2)-i}-x_{h(2)}$ and $y_{h(2)-i}-y_{h(2)}$ as $\xi_{2i}^s$ and $\xi_{2i+1}^s$, respectively, the tile shape $U$ is then parameterized as follows (only the $xy$-coordinates of the first and second tiling edges are shown):
\begin{equation}
\label{eq:parameterize_u_IH7}
%\footnotesize
\thickmuskip=0mu
%\medmuskip=0mu
\thinmuskip=0mu
\renewcommand{\arraystretch}{1.20}
\begin{bmatrix*}[c]
\colorbox[gray]{.9}{$x_{h(1)}$} \\
x_{h(2)-k_1} \\
\scalebox{1.0}[0.8]{\vdots} \\
x_{h(2)-1} \\
\colorbox[gray]{.9}{$x_{h(2)}$} \\
x_{h(2)+1} \\
\scalebox{1.0}[0.8]{\vdots} \\
x_{h(2)+k_1} \\
\colorbox[gray]{.9}{$x_{h(3)}$} \\
\scalebox{1.0}[0.8]{\vdots} \\
\colorbox[gray]{.9}{$y_{h(1)}$} \\
y_{h(2)-k_1} \\
\scalebox{1.0}[0.8]{\vdots} \\
y_{h(2)-1} \\
\colorbox[gray]{.9}{$y_{h(2)}$} \\
y_{h(2)+1} \\
\scalebox{1.0}[0.8]{\vdots} \\
y_{h(2)+k_1} \\
\colorbox[gray]{.9}{$y_{h(3)}$} \\
\scalebox{1.0}[0.8]{\vdots} \\
\end{bmatrix*} 
=
\renewcommand{\arraystretch}{1.18}
\begin{bmatrix*}[c]
\colorbox[gray]{.9}{$\bm{d_1}^{\top}$} \\
\bm{d_2}^{\top} \\
\scalebox{1.0}[0.8]{\vdots} \\
\bm{d_2}^{\top} \\
\colorbox[gray]{.9}{$\bm{d_2}^{\top}$} \\
\bm{d_2}^{\top} \\
\scalebox{1.0}[0.8]{\vdots} \\
\bm{d_2}^{\top} \\
\colorbox[gray]{.9}{$\bm{d_3}^{\top}$} \\
\scalebox{1.0}[0.8]{\vdots} \\
\colorbox[gray]{.9}{$\bm{d_{n_v+1}}^{\top}$} \\
\bm{d_{n_v+2}}^{\top} \\
\scalebox{1.0}[0.8]{\vdots} \\
\bm{d_{n_v+2}}^{\top} \\
\colorbox[gray]{.9}{$\bm{d_{n_v+2}}^{\top}$} \\
\bm{d_{n_v+2}}^{\top} \\
\scalebox{1.0}[0.8]{\vdots} \\
\bm{d_{n_v+2}}^{\top} \\
\colorbox[gray]{.9}{$\bm{d_{n_v+3}}^{\top}$} \\
\scalebox{1.0}[0.8]{\vdots} \\
\end{bmatrix*}
\bm{\xi_d}
+
\renewcommand{\arraystretch}{1.25}
\tfrac{\xi_1^s}{\sqrt{2}}
\begin{bmatrix*}[c]
\colorbox[gray]{.9}{$\color[gray]{0.9}{{}^1 }$} \\
\\
 \\
1 \\
\colorbox[gray]{.9}{$\color[gray]{0.9}{{}^1 }$} \\
\cos \theta \\
 \\
\\
\colorbox[gray]{.9}{$\color[gray]{0.9}{{}^1 }$} \\
 \\
\colorbox[gray]{.9}{$\color[gray]{0.9}{{}^1 }$} \\
\\
 \\
0 \\
\colorbox[gray]{.9}{$\color[gray]{0.9}{{}^1 }$} \\
\sin \theta \\
 \\
\\
\colorbox[gray]{.9}{$\color[gray]{0.9}{{}^1 }$} \\
\scalebox{1.0}[0.8]{\vdots} \\
\end{bmatrix*}
+
\tfrac{\xi_{k_2}^s}{\sqrt{2}}
\begin{bmatrix*}[c]
\colorbox[gray]{.9}{$\color[gray]{0.9}{{}^1 }$} \\
\\
 \\
0 \\
\colorbox[gray]{.9}{$\color[gray]{0.9}{{}^1 }$} \\
-\sin \theta \\
 \\
\\
\colorbox[gray]{.9}{$\color[gray]{0.9}{{}^1 }$} \\
 \\
\colorbox[gray]{.9}{$\color[gray]{0.9}{{}^1 }$} \\
\\
 \\
1 \\
\colorbox[gray]{.9}{$\color[gray]{0.9}{{}^1 }$} \\
\cos \theta \\
 \\
\\
\colorbox[gray]{.9}{$\color[gray]{0.9}{{}^1 }$} \\
\scalebox{1.0}[0.8]{\vdots} \\
\end{bmatrix*}
+
\cdots
+
\tfrac{\xi_{2 k_1}^s}{\sqrt{2}}
\begin{bmatrix*}[c]
\colorbox[gray]{.9}{$\color[gray]{0.9}{{}^1 }$} \\
1 \\
 \\
\\
\colorbox[gray]{.9}{$\color[gray]{0.9}{{}^1 }$} \\
\\
 \\
\cos \theta \\
\colorbox[gray]{.9}{$\color[gray]{0.9}{{}^1 }$} \\
 \\
\colorbox[gray]{.9}{$\color[gray]{0.9}{{}^1 }$} \\
0 \\
\\
\\
\colorbox[gray]{.9}{$\color[gray]{0.9}{{}^1 }$} \\
\\
 \\
\sin \theta \\
\colorbox[gray]{.9}{$\color[gray]{0.9}{{}^1 }$} \\
\scalebox{1.0}[0.8]{\vdots} \\
\end{bmatrix*}
+
\tfrac{\xi_{2 k_1+1}^s}{\sqrt{2}}
\begin{bmatrix*}[c]
\colorbox[gray]{.9}{$\color[gray]{0.9}{{}^1 }$} \\
0 \\
 \\
\\
\colorbox[gray]{.9}{$\color[gray]{0.9}{{}^1 }$}\\
\\
 \\
-\sin \theta \\
\colorbox[gray]{.9}{$\color[gray]{0.9}{{}^1 }$} \\
 \\
\colorbox[gray]{.9}{$\color[gray]{0.9}{{}^1 }$} \\
1 \\
 \\
\\
\colorbox[gray]{.9}{$\color[gray]{0.9}{{}^1 }$} \\
\\
 \\
\cos \theta \\
\colorbox[gray]{.9}{$\color[gray]{0.9}{{}^1 }$} \\
\scalebox{1.0}[0.8]{\vdots} \\
\end{bmatrix*}
%+
\cdots
,
\end{equation}
where the blank elements in the column vectors are zero. The matrix $B'_d$ (see Eq.~(\ref{eq:B'd_IH51})) is then given by the following formula:
\begin{equation} 
%\footnotesize
\renewcommand{\arraystretch}{1.15}
\label{eq:B'd_IH7}
{B'}_d =
\begin{array}{rccll}
\ldelim[{21.5}{2pt}[] & & \cellcolor[gray]{.9}\bm{d_1}^{\top}  & \rdelim]{21.5}{2pt}[] & :h(1) \\
% & &  &  &  \\
                    & & {\bm{d^b_2}}^{\top} &   & \rdelim\}{3}{10pt}[$h(2)-i \ (i = 1, \dots, k_1)$] \\ 
                    & & \scalebox{1.0}[0.8]{\vdots} &   & \\                  
                    & & {\bm{d^b_2}}^{\top} &   & \\
                    & & \cellcolor[gray]{.9}\bm{d_2}^{\top}     &   & :h(2)  \\ 
                    & & {\bm{d^f_2}}^{\top} &   & \rdelim\}{3}{10pt}[$h(2)+i \ (i = 1, \dots, k_1)$] \\ 
                    & & \scalebox{1.0}[0.8]{\vdots}  &   & \\                  
                    & & {\bm{d^f_2}}^{\top} &   & \\
                    & & \cellcolor[gray]{.9}{\bm{d_3}}^{\top}   &   & :h(3)  \\ 
                    & & \scalebox{1.0}[0.8]{\vdots}  &   & \\
                    & & \cellcolor[gray]{.9}{\bm{d_{n_v+1}}}^{\top} &   & :h(1) \\ 
                    & & {\bm{d^b_{n_v+2}}}^{\top}  &  & \rdelim\}{3}{10pt}[$h(2)-i \ (i = 1, \dots, k_1)$] \\ 
                    & & \scalebox{1.0}[0.8]{\vdots} &   & \\                  
                    & & {\bm{d^b_{n_v+2}}}^{\top} &  & \\ 
                    & & \cellcolor[gray]{.9}{\bm{d_{n_v+2}}}^{\top} &  & :h(2) \\  
                    & & {\bm{d^f_{n_v+2}}}^{\top}  &  & \rdelim\}{3}{10pt}[$h(2)+i \ (i = 1, \dots, k_1)$] \\ 
                    & & \scalebox{1.0}[0.8]{\vdots} &   & \\
                    & & {\bm{d^f_{n_v+2}}}^{\top}  &  & \\
                    & & \cellcolor[gray]{.9}{\bm{d_{n_v+3}}}^{\top}   &   & :h(3) \\ 
                    & & \scalebox{1.0}[0.8]{\vdots}  &   & \\
\end{array}
,
\vspace{7mm}
\end{equation} 
where the row vectors of $B'_d$ are given by 
\begin{equation} 
%\footnotesize
\thickmuskip=0mu
%\medmuskip=0mu
\thinmuskip=0mu
\begin{bmatrix*}[c]
{\bm{d^b_2}}^{\top} \\ 
{\bm{d^f_2}}^{\top} \\
{\bm{d^b_{n_v+2}}}^{\top} \\ 
{\bm{d^f_{n_v+2}}}^{\top} 
\end{bmatrix*}  
= 
\renewcommand{\arraystretch}{1.3}
\begin{bmatrix*}[c]
{\bm{d_2}}^{\top} \\ 
{\bm{d_2}}^{\top} \\
{\bm{d_{n_v+2}}}^{\top} \\ 
{\bm{d_{n_v+2}}}^{\top} 
\end{bmatrix*}  
-
\dfrac{1}{2}
{
\begin{bmatrix*}[c]
1 \\
\cos \theta \\
0 \\ 
\sin \theta 
\end{bmatrix*}  
}^{\top}
\begin{bmatrix*}[c]
{\bm{d_2}}^{\top} \\ 
{\bm{d_2}}^{\top} \\
{\bm{d_{n_v+2}}}^{\top} \\ 
{\bm{d_{n_v+2}}}^{\top} 
\end{bmatrix*}  
\begin{bmatrix*}[c]
1 \\
\cos \theta \\
0 \\ 
\sin \theta 
\end{bmatrix*}  
-
\dfrac{1}{2}
{
\begin{bmatrix*}[c]
0 \\
-\sin \theta \\
1 \\ 
\cos \theta 
\end{bmatrix*}  
}^{\top}
\begin{bmatrix*}[c]
{\bm{d_2}}^{\top} \\ 
{\bm{d_2}}^{\top} \\
{\bm{d_{n_v+2}}}^{\top} \\ 
{\bm{d_{n_v+2}}}^{\top} 
\end{bmatrix*}  
\begin{bmatrix*}[c]
0 \\
-\sin \theta \\
1 \\ 
\cos \theta 
\end{bmatrix*}  
.
\end{equation} 

Obviously, we can construct the matrix $B_{ik}$ in $O(n)$ time, because the matrix $B_s$ is sparse, all column vectors of $B_s$ are mutually orthogonal, and the matrix $B'_d$ is obtained in $O(n)$ time.

In fact, we must consider $\theta=-120^{\circ}$ as well in the template of IH7. This can be easily implemented by considering the reverse numbering scheme of the goal polygon. The same process is necessary for the isohedral types IH21 and IH28.

%%%%%%%%%%%%%%%%%%%%%%%%%%%%%%%%%% supplementary material %%%%%%%%%%%%%%%%%%%%%%%%%%%%%%%%%
\newpage
\setcounter{figure}{0}
\setcounter{footnote}{0}
\setcounter{part}{0}

\begin{center}

\begin{spacing}{1.7}
{\LARGE {Supplementary file to ``An Efficient Algorithm for the Escherization Problem in the Polygon Representation''}}\\[12pt]
\end{spacing}
% Authors and addresses:

\footnotesize

\mbox{\large Yuichi Nagata$^{\ast}$, Shingi Imahori$^\dagger$}\\[8pt]
${}^{\ast}$Department of Science and Technology, Tokushima University, 2-1 Minami-jousanjima, Tokushima-shi, Tokushima 770-8506, Japan\\[6pt]
${}^{\dagger}$Department of Information and System Engineering Faculty of Science and Engineering, Chuo University, Bunkyo-ku, Tokyo 112-8551, Japan\\[6pt]
\normalsize

\end{center}

% begin "double spacing" the text:
\baselineskip 15.3pt plus .3pt minus .1pt

% Here is the abstract:

\noindent\hrulefill

\begin{figure} [t] %%%%%%%%%%%%%%%%%%%%%%%%%%%%%
\centering
\includegraphics[scale=0.28,keepaspectratio,clip]{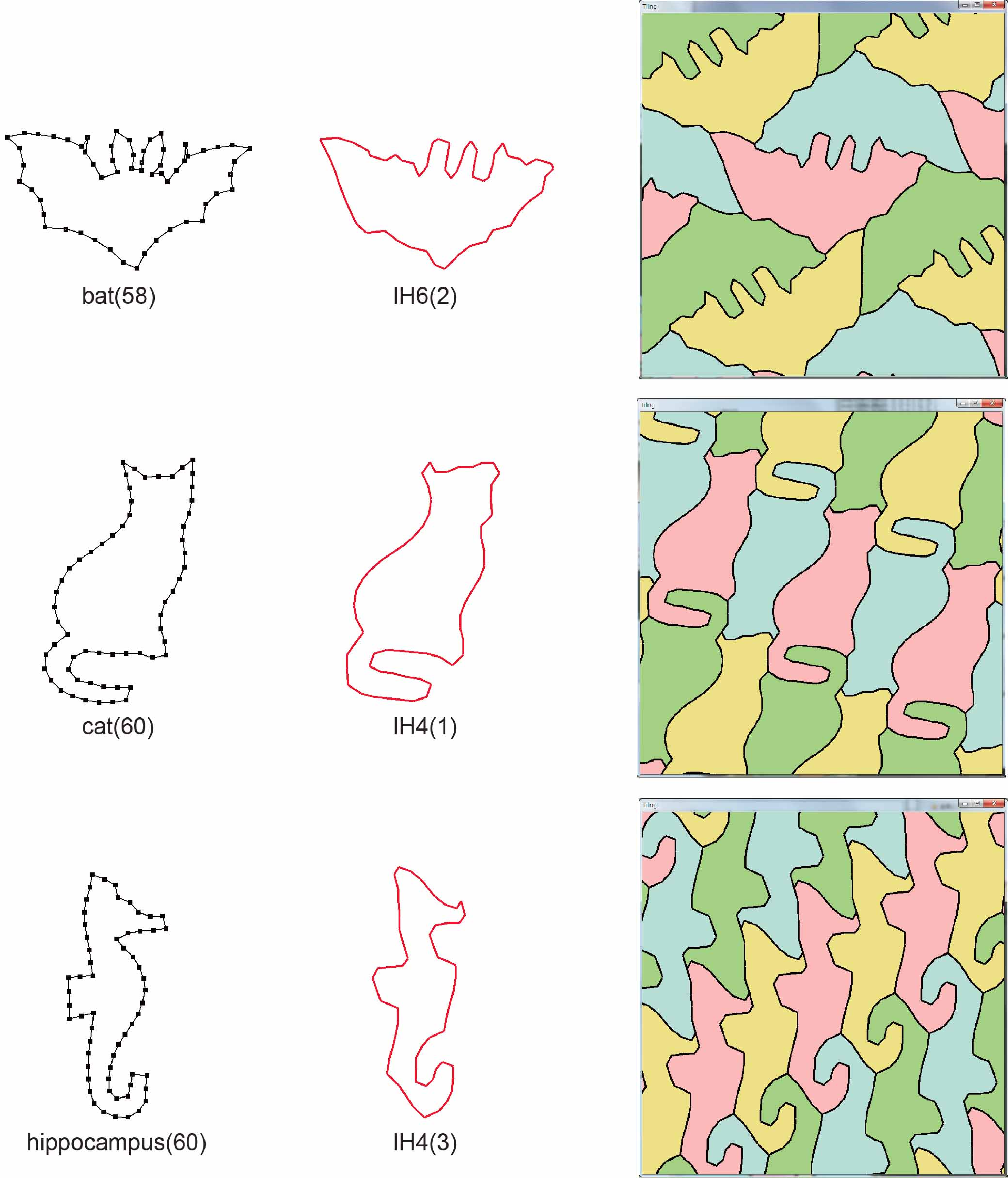}
\caption{Goal polygons (left), most satisfactory tile shapes (middle), and tilings (right) for 24 goal shapes.}
\label{fig:tile_shape_all}
\end{figure}

We applied the proposed algorithm (Algorithm 1 that incorporates the mechanism to evaluate a lower bound on the value of $eval_{ikj}$ to eliminate unnecessary computation of $eval_{ikj}$ that do not yield an improvement in $eval_{mim}$) to 24 goal shapes (polygons) shown in the left side of Figure \ref{fig:tile_shape_all}. However, the optimal tile shape may not be the most satisfactory shape because the distance metric (the Procrustes distance or Euclidean distance) used to measure the similarity between two polygons does not necessarily coincide with human intuition. Therefore, we checked the top 20 tile shapes\footnote{In order to do this, we need to modify Algorithm 1 so that $eval_{min}$ represents the 20th minimum value of $eval_{ikj}$ found so far.} among all combinations $i \in I$, $k \in K_i$, and $j \in J$, and then selected the most satisfactory one.

We present the selected tile shapes in the middle of the figure along with the corresponding isohedral types, where the numbers in parentheses indicate the overall ranking of the similarity in terms of the Procrustes (Euclidean) distance. Tilings generated from these tile shapes are shown in the right side of the figure. 

 %%%%%%%%%%%%%%%%%%%%%%%%%%%%%

\addtocounter{figure}{-1}
\begin{figure} [t] %%%%%%%%%%%%%%%%%%%%%%%%%%%%%
\centering
\includegraphics[scale=0.28,keepaspectratio,clip]{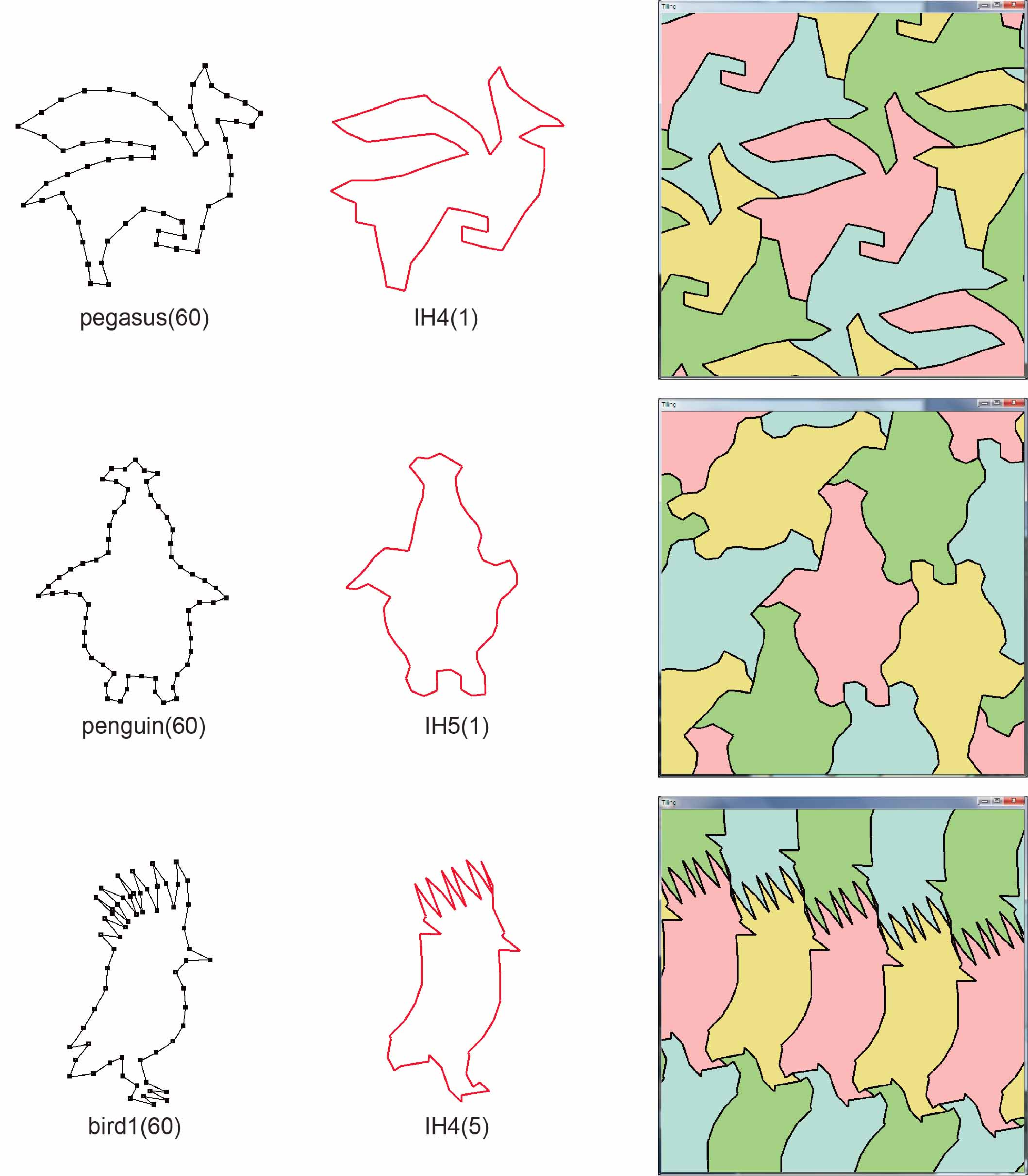}
\caption{Goal polygons (left), most satisfactory tile shapes (middle), and tilings (right) for 24 goal shapes. (continued)}
%\label{fig:tile_shape_all}
\end{figure} %%%%%%%%%%%%%%%%%%%%%%%%%%%%%

\addtocounter{figure}{-1}
\begin{figure} [t] %%%%%%%%%%%%%%%%%%%%%%%%%%%%%
\centering
\includegraphics[scale=0.28,keepaspectratio,clip]{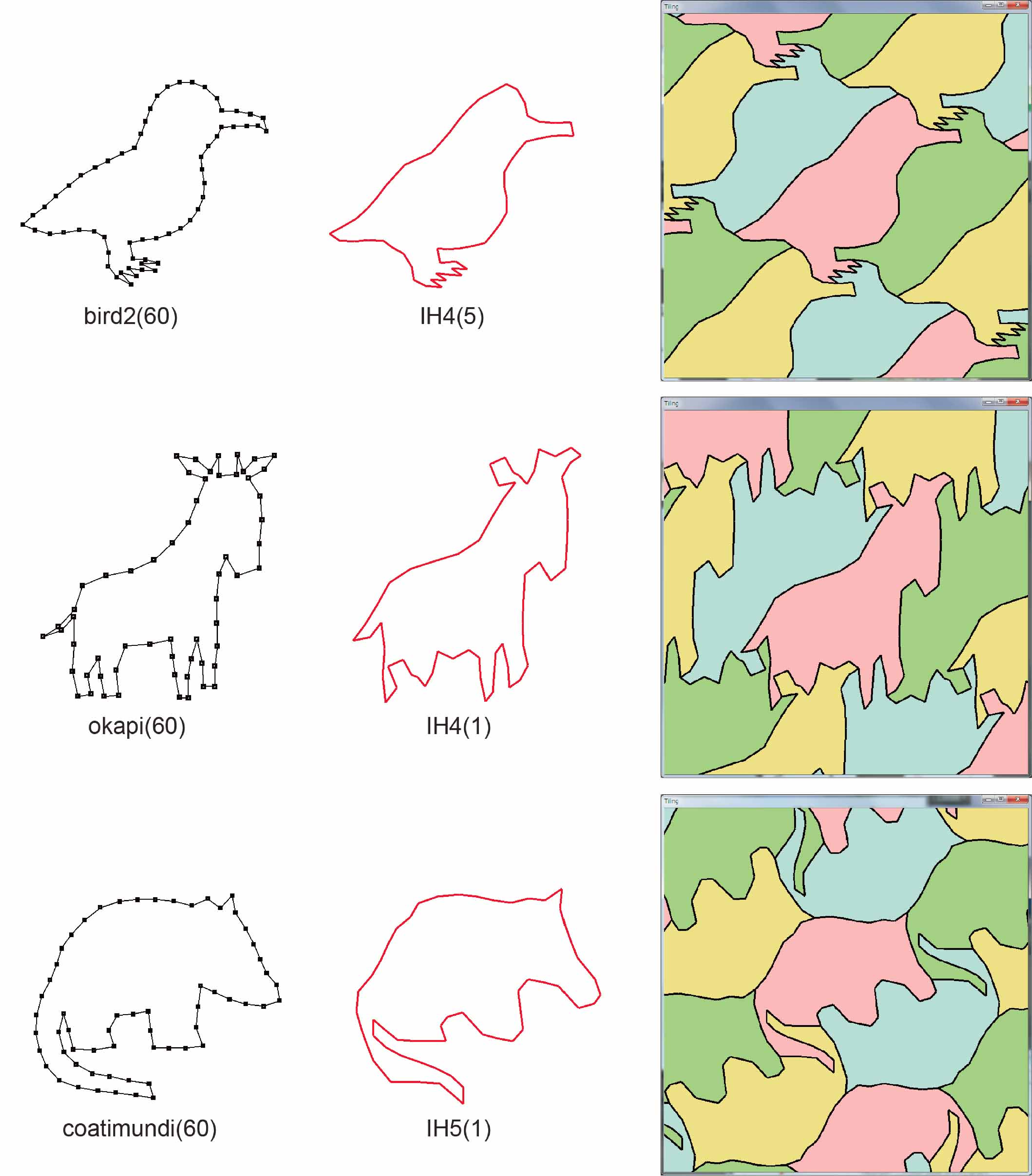}
\caption{Goal polygons (left), most satisfactory tile shapes (middle), and tilings (right) for 24 goal shapes. (continued)}
%\label{fig:tile_shape_all}
\end{figure} %%%%%%%%%%%%%%%%%%%%%%%%%%%%%

\addtocounter{figure}{-1}
\begin{figure} [t] %%%%%%%%%%%%%%%%%%%%%%%%%%%%%
\centering
\includegraphics[scale=0.28,keepaspectratio,clip]{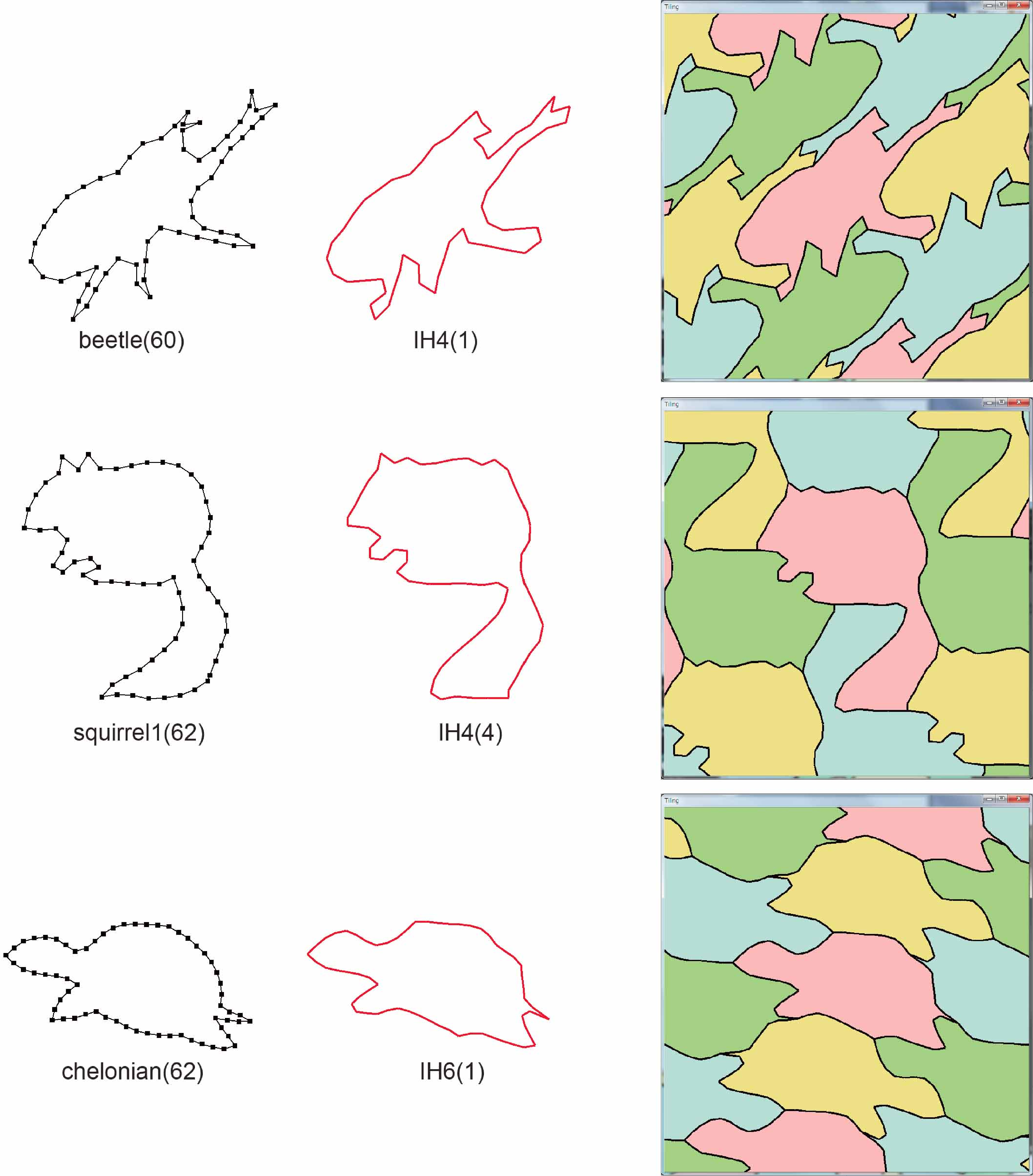}
\caption{Goal polygons (left), most satisfactory tile shapes (middle), and tilings (right) for 24 goal shapes. (continued)}
%\label{fig:tile_shape_all}
\end{figure} %%%%%%%%%%%%%%%%%%%%%%%%%%%%%

\addtocounter{figure}{-1}
\begin{figure} [t] %%%%%%%%%%%%%%%%%%%%%%%%%%%%%
\centering
\includegraphics[scale=0.28,keepaspectratio,clip]{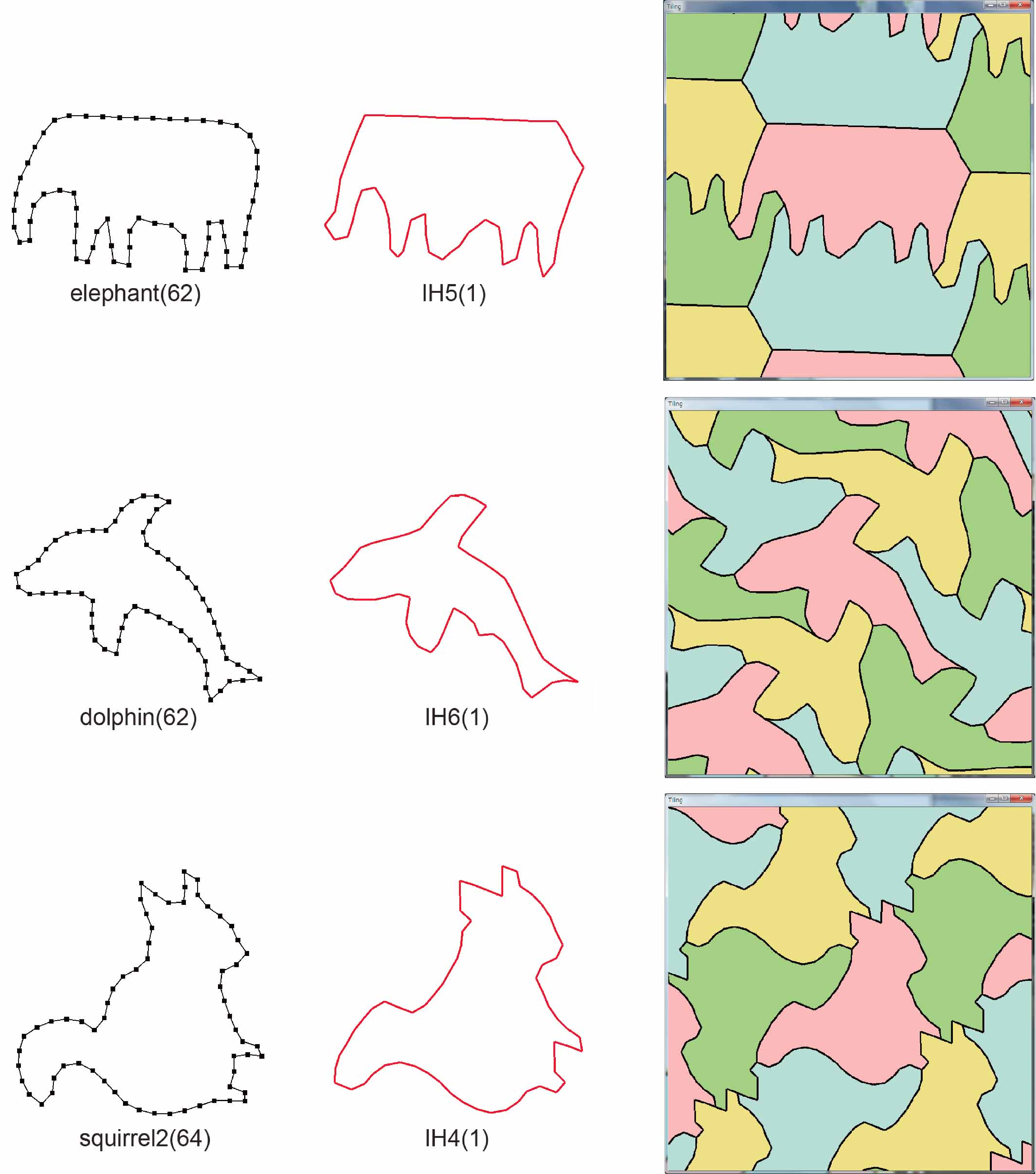}
\caption{Goal polygons (left), most satisfactory tile shapes (middle), and tilings (right) for 24 goal shapes. (continued)}
%\label{fig:tile_shape_all}
\end{figure} %%%%%%%%%%%%%%%%%%%%%%%%%%%%%

\addtocounter{figure}{-1}
\begin{figure} [t] %%%%%%%%%%%%%%%%%%%%%%%%%%%%%
\centering
\includegraphics[scale=0.28,keepaspectratio,clip]{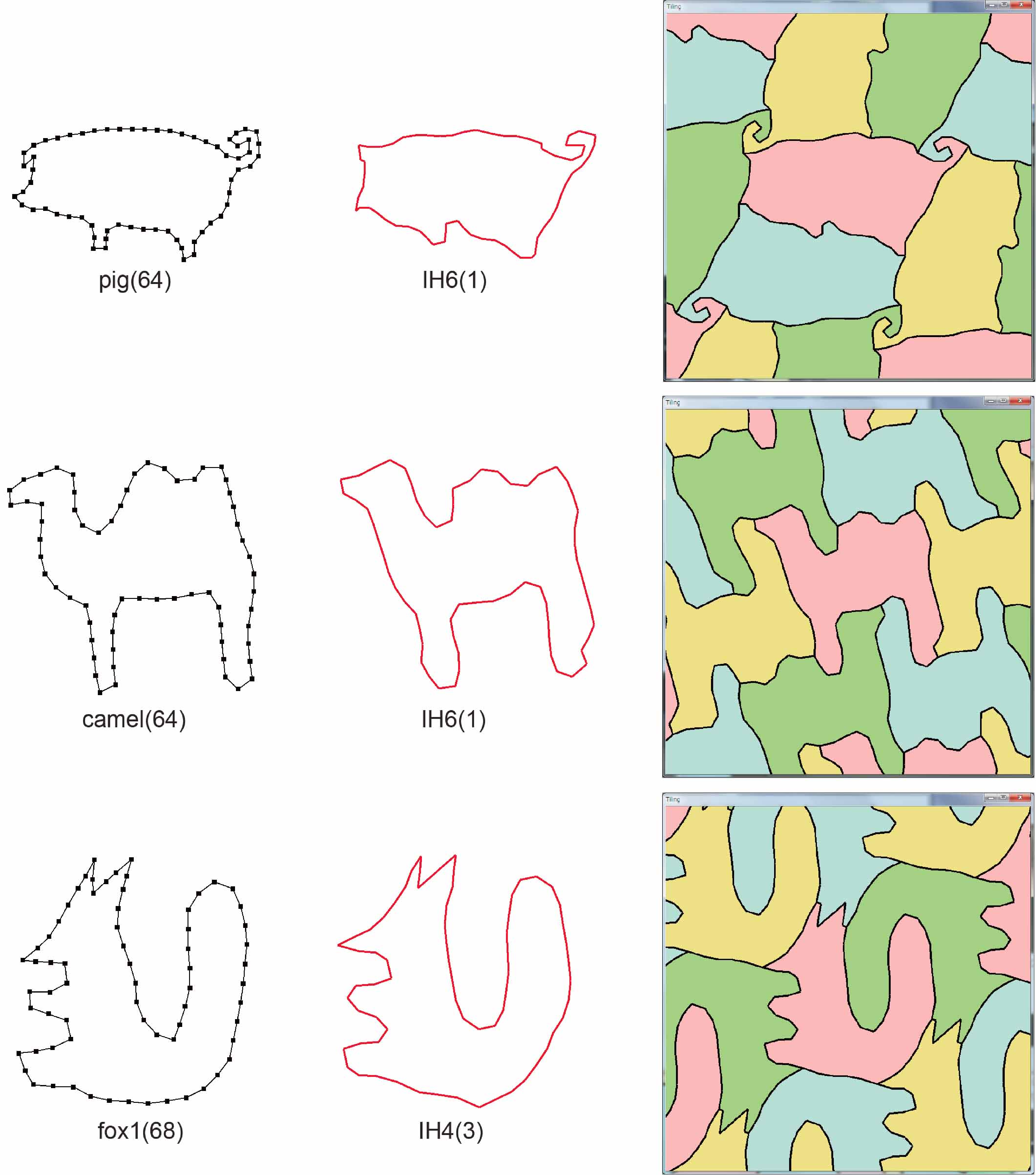}
\caption{Goal polygons (left), most satisfactory tile shapes (middle), and tilings (right) for 24 goal shapes. (continued)}
%\label{fig:tile_shape_all}
\end{figure} %%%%%%%%%%%%%%%%%%%%%%%%%%%%%

\addtocounter{figure}{-1}
\begin{figure} [t] %%%%%%%%%%%%%%%%%%%%%%%%%%%%%
\centering
\includegraphics[scale=0.28,keepaspectratio,clip]{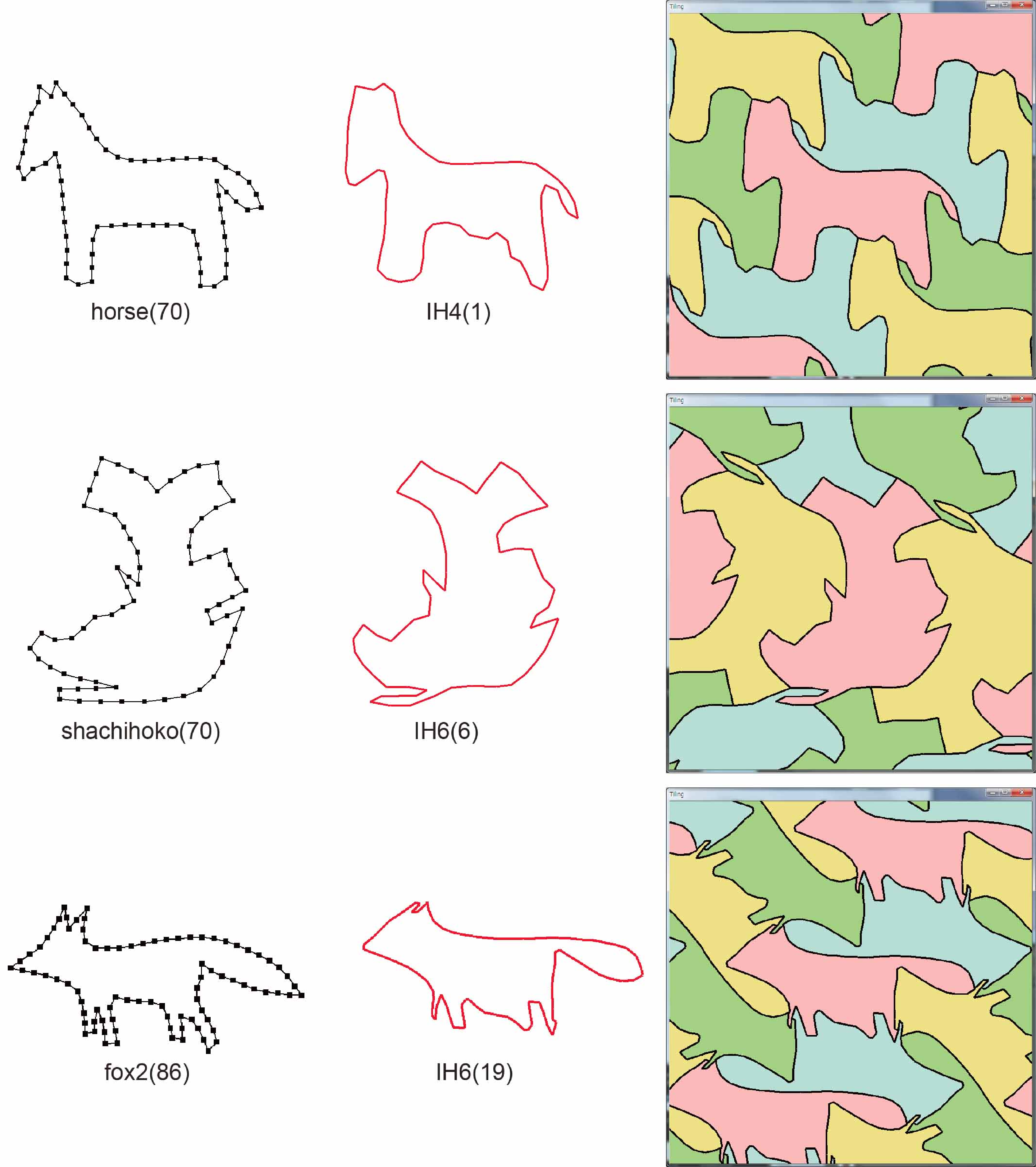}
\caption{Goal polygons (left), most satisfactory tile shapes (middle), and tilings (right) for 24 goal shapes. (continued)}
%\label{fig:tile_shape_all}
\end{figure} %%%%%%%%%%%%%%%%%%%%%%%%%%%%%

\addtocounter{figure}{-1}
\begin{figure} [t] %%%%%%%%%%%%%%%%%%%%%%%%%%%%%
\centering
\includegraphics[scale=0.28,keepaspectratio,clip]{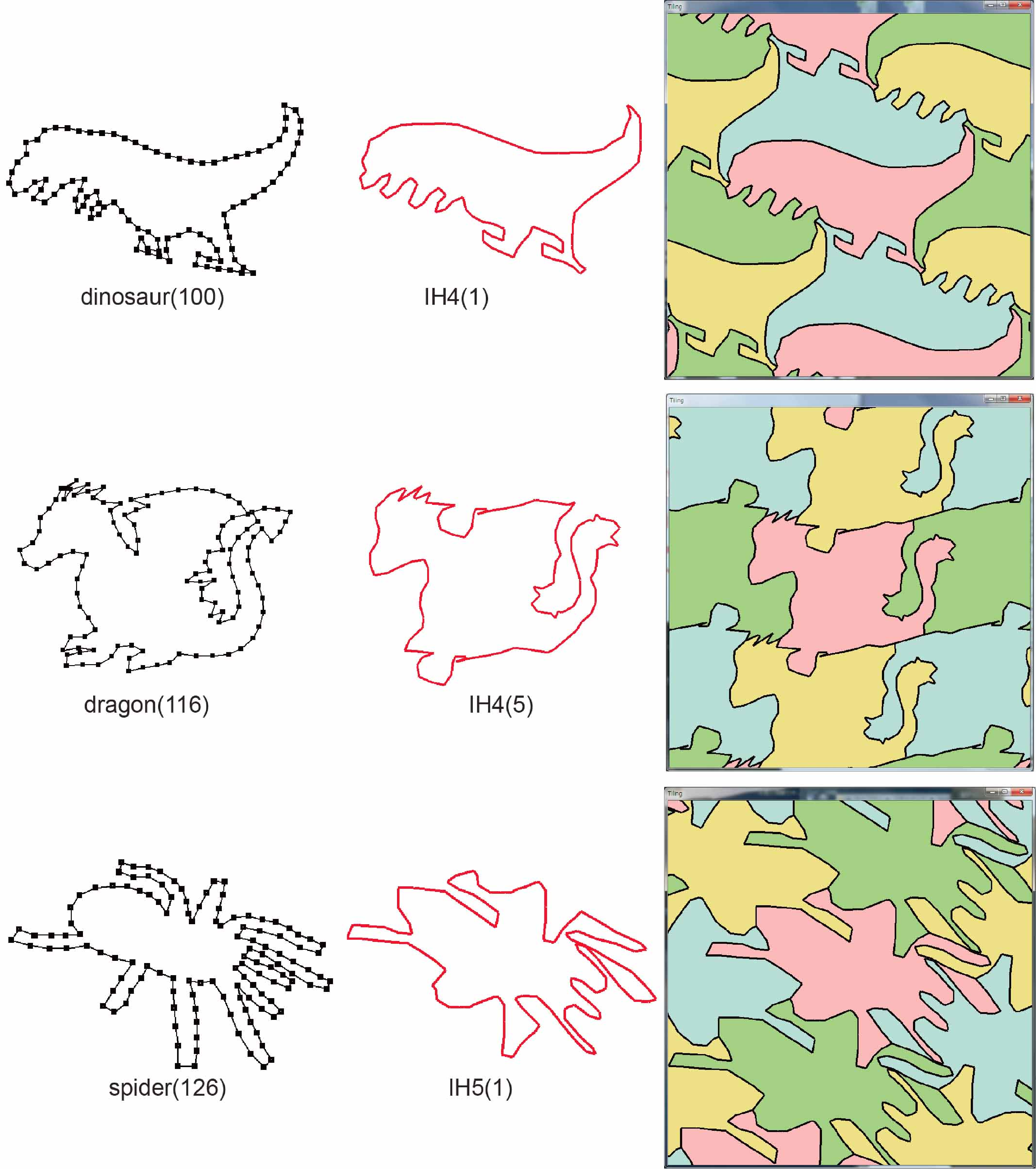}
\caption{Goal polygons (left), most sativsfactory tile shapes (middle), and tilings (right) for 24 goal shapes. (continued)}
%\label{fig:tile_shape_all}
\end{figure} %%%%%%%%%%%%%%%%%%%%%%%%%%%%%

\end{document}